\documentclass[manuscript=article]{achemso}
\usepackage{natbib}
\usepackage{xr}
\makeatletter
\newcommand*{\addFileDependency}[1]{
  \typeout{(#1)}
  \@addtofilelist{#1}
  \IfFileExists{#1}{}{\typeout{No file #1.}}
}
\makeatother

\usepackage[version=3]{mhchem} 
\usepackage{siunitx}
\usepackage{amssymb}
\usepackage{amsmath}
\usepackage{lipsum}
\usepackage{multicol}
\usepackage[labelfont=bf]{caption} 
\usepackage{booktabs,caption}
\usepackage{threeparttable}
\usepackage{titlesec,hyperref}
\usepackage{nameref}
\usepackage{color}
\usepackage{sidecap, caption}
\usepackage{chemfig}
\usepackage{adjustbox}
\usepackage{graphicx}
\usepackage{placeins}
\usepackage{chemfig}
\usepackage{array}
\usepackage{comment}
\usepackage{tabularx,booktabs}
\DeclareSIUnit\bar{bar}

\newcommand{\beginsupplement}{%
        \setcounter{table}{0}
        \renewcommand{\thetable}{S\arabic{table}}%
        \setcounter{figure}{0}
        \renewcommand{\thefigure}{S\arabic{figure}}%
        \setcounter{equation}{0}
        \renewcommand{\theequation}{S\arabic{equation}}%
     }

\newcommand{\redline}{\raisebox{2pt}{\protect\tikz{\protect\draw[-,red,solid,line width = 0.9pt](0,0) -- (5mm,0);}}}
\newcommand{\reddotline}{\raisebox{2pt}{\protect\tikz{\protect\draw[-,red,dashed,line width = 0.9pt](0,0) -- (5mm,0);}}}
\newcommand{\bluedotline}{\raisebox{2pt}{\protect\tikz{\protect\draw[-,blue,dashed,line width = 0.9pt](0,0) -- (5mm,0);}}}
\newcommand{\blackdotline}{\raisebox{2pt}{\protect\tikz{\protect\draw[-,black,dashed,line width = 0.9pt](0,0) -- (5mm,0);}}}

\newcommand{\bluestar}{\raisebox{0pt}{\protect\tikz{\protect\filldraw[fill=blue, draw=blue] (0,0) -- (0.35mm,1mm) -- (1.5mm,1mm) -- (0.55mm,1.5mm) -- (0.9mm,2.5mm) -- (0,2mm) -- (-0.9mm,2.5mm) -- (-0.55mm,1.5mm) -- (-1.5mm,1mm) -- (-0.35mm,1mm) -- cycle;}}}
\newcommand{\redstar}{\raisebox{0pt}{\protect\tikz{\protect\filldraw[fill=red, draw=red] (0,0) -- (0.35mm,1mm) -- (1.5mm,1mm) -- (0.55mm,1.5mm) -- (0.9mm,2.5mm) -- (0,2mm) -- (-0.9mm,2.5mm) -- (-0.55mm,1.5mm) -- (-1.5mm,1mm) -- (-0.35mm,1mm) -- cycle;}}}

\newcommand{\bluewithcircle}{\raisebox{0pt}{\protect\tikz{\protect\filldraw[fill=blue, draw=blue](2.5mm,0) circle [radius=0.8mm];\protect\draw[-,blue,solid,line width = 1.0pt](0,0) -- (1.7mm,0);\protect\draw[-,blue,solid,line width = 1.0pt](3.3mm,0) -- (5mm,0);}}}

\newcommand{\orangewithcircle}{\raisebox{0pt}{\protect\tikz{\protect\filldraw[fill=orange, draw=orange](2.5mm,0) circle [radius=0.9mm];\protect\draw[-,orange,solid,line width = 1.0pt](0,0) -- (5mm,0);}}}

\newcommand{\orangediamond}{\raisebox{0pt}{\protect\tikz{\protect\filldraw[fill=orange, draw=orange] (0,0) -- (1.5mm,1.5mm) -- (0,3mm) -- (-1.5mm,1.5mm) -- cycle;}}}

\newcommand{\greencircle}{\raisebox{0pt}{\protect\tikz{\protect\filldraw[fill=green, draw=green](2.5mm,0) circle [radius=1mm];\protect\filldraw[fill=green, draw=green](2.5mm,0) circle [radius=0.2mm];}}}

\newcommand{\greentriiangle}{\raisebox{0pt}{\protect\tikz{\protect\filldraw[fill=green, draw=green](2.5mm,0) (0.15,-0.1) -- (0.35,-0.1) -- (0.25,0.1)-- (0.15,-0.1);}}}

\newcommand{\redtriiangle}{\raisebox{0pt}{\protect\tikz{\protect\filldraw[fill=red, draw=red](2.5mm,0) (0.15,-0.1) -- (0.35,-0.1) -- (0.25,0.1)-- (0.15,-0.1);}}}

\newcommand{\orangetriiangle}{\raisebox{0pt}{\protect\tikz{\protect\filldraw[fill=orange, draw=orange](2.5mm,0) (0.15,-0.1) -- (0.35,-0.1) -- (0.25,0.1)-- (0.15,-0.1);}}}

\newcommand{\orangewithsquare}{\raisebox{0pt}{\protect\tikz{\protect\filldraw[fill=orange, draw=orange](1.75mm, -0.75mm) rectangle ++(4.5pt,4.5pt);\protect\draw[orange,solid](0,0) -- (5mm,0);}}}
\newcommand{\blackmultiline}{\raisebox{0pt}{\protect\tikz{\protect\draw[thick, black] (-0.1,-0.1) -- (0.1,0.1) (-0.1,0.1) -- (0.1,-0.1) (-0.15,0) -- (0.15,0);}}}

\newcommand{\pinkplus}{\raisebox{0pt}{\protect\tikz{\protect\draw[thick, pink] (-0.1,0) -- (0.1,0) (0,-0.1) -- (0,0.1);}}}
\newcommand{\oliveinvertedtriangle}{\raisebox{0pt}{\protect\tikz{\protect\filldraw[fill=olive, draw=olive](0,-1mm) -- (-1mm,1mm) -- (1mm,1mm) -- cycle;}}}

\newcommand{\reddiamond}{\raisebox{0pt}{\protect\tikz{\protect\filldraw[fill=red, draw=red, rotate=45](1.75mm, 0.75mm) rectangle ++(4.5pt,4.5pt);}}}
\newcommand{\graystar}{\raisebox{0pt}{\protect\tikz{\protect\filldraw[fill=gray, draw=gray] (0,0) -- (0.35mm,1mm) -- (1.5mm,1mm) -- (0.55mm,1.5mm) -- (0.9mm,2.5mm) -- (0,2mm) -- (-0.9mm,2.5mm) -- (-0.55mm,1.5mm) -- (-1.5mm,1mm) -- (-0.35mm,1mm) -- cycle;}}}
\newcommand{\bluewithsquare}{\raisebox{0pt}{\protect\tikz{\protect\filldraw[fill=blue, draw=blue](1.75mm, -0.75mm) rectangle ++(4.5pt,4.5pt);\protect\draw[blue,solid](0,0) -- (5mm,0);}}}
\newcommand{\yellowmultiline}{\raisebox{0pt}{\protect\tikz{\protect\draw[thick, yellow] (-0.1,-0.1) -- (0.1,0.1) (-0.1,0.1) -- (0.1,-0.1) (-0.15,0) -- (0.15,0);}}}

\newcommand{\redwithsquare}{\raisebox{0pt}{\protect\tikz{\protect\filldraw[fill=white, draw=red](1.75mm, -0.75mm) rectangle ++(4.5pt,4.5pt);\protect\draw[red,solid](0,0) -- (5mm,0);}}}

\author{Ankit Patidar}
\email{a786.pati@gmail.com}

\author{Gaurav Goel}
\affiliation{Department of Chemical Engineering, Indian Institute of Technology Delhi, City - New Delhi, Postal code - 110016, India}
\email{goelg@chemical.iitd.ac.in}
\phone{91 11 2659 1025}

\title[An \textsf{achemso} demo]
  {Characterization and Quantification of Immiscible Polymer Blend Compatibilization by Phyllosilicate Clays}
\abbreviations{AA, CG, TMA-MMT, TPS, $R_\mathrm{g}$, $S_\mathrm{c}$, $S_2$, RDF, MSD, $Ar$, $Ln$, $\rho(z)$}
\keywords{American Chemical Society, \LaTeX}

\begin{document}

\begin{tocentry}
{\centering
}
\end{tocentry}
\mciteErrorOnUnknownfalse
\begin{abstract}
Phyllosilicate clays have significant applications in various polymer nanocomposites due to their high anisotropy and tunable surface polarity. Their distribution and interface localization in polymer blends can be used to tune the properties of polymer-clay nanocomposites (PCNCs). A coarse-grained (CG) force field for the clay particles can significantly aid applications of molecular simulations in PCNC development. Here, we developed MARTINI-3 parameters, a CG force field with high chemical specificity, for phyllosilicate clays with diverse surface polarities. An initial interaction level for the clay particle functional groups was determined from hydration free energies, obtained by applying the Lifshitz theory to experimental surface tension data. These were fine-tuned using the structural, thermodynamic, and dynamic properties of thermoplastic starch (TPS)-clay composites obtained from all-atom (AA) molecular dynamics (MD) simulations. An accurate estimate was obtained for the radial distribution function and two-body excess entropy for various TPS components, properties not used in CG parameterization, establishing the robustness of the developed CG parameters. We investigated the effect of clay surface polarity on polymer segmental dynamics and the structure-property relationships in these systems. The developed CG parameters were used to study the effect of dodecyltrimethylammonium (C12TAB)-modified montmorillonite (MMT), an organically-modified MMT, on TPS-polyethylene (PE) blend morphology using large time- ($\sim$ $10^3$ \SI{}{\nano\second}) and large lengthscale ($\sim$ $10^2$ \SI{}{\nano\meter}) MD simulations. We observed effective compatibilization of the TPS-PE interphase by the amphiphillic clay particle, leading to a reduction in the TPS-PE interfacial tension from \SI{45}{\milli\newton\per\meter} to \SI{13.06}{\milli\newton\per\meter}. 
We found a high qualitative and quantitative agreement between MARTINI-3 estimates for distinct properties of model MMT-based PCNCs and those from AA simulations and available experimental data, establishing grounds for the transferability of developed parameters to other systems of interest.
\end{abstract}
\maketitle

\section{Introduction}
The environmental impact of synthetic materials like polyethylene (PE) in packaging waste has driven significant research into biopolymers \cite{Marsh2007}. Thermoplastic starch (TPS) can be used as an alternative due to its biodegradability \cite{babu2013current}, low cost, wide availability, and excellent film-forming properties (good transparency and low processing temperature) \cite{jimenez2012edible,attaran2017materials}. TPS formulations consisting of linear amylose (1-4 linked $\alpha$-D glucose), amylopectin (backbone of 1-4 linked $\alpha$-D glucose with additional 1-6 linked branches), and plasticizers (sorbitol, water, etc.) \cite{carvalho2008starch}, have been shown as a suitable option for some food packaging applications, either as films or as a component in paper coating additives \cite{khan2017thermoplastic, stepto2003processing, de2022thermoplastic,kaboorani2021tailoring, correa2022biodegradability}. However, its poor mechanical properties (low Young's modulus and tensile strength, limited elongation at break) \cite{Cyras2008,chocyk2015influence,domene2019influence} and low moisture resistance \cite{salaberria2015role} necessitate the addition of components with complementary properties, such as addition of nanofillers \cite{Rezaei2015, zhou2015starch, Suter2015} and blending with hydrophobic polymers like PE \cite{altayan2022toward,baumberger1998use}. PE-TPS formulations were shown to have good film-forming properties with the additional advantage of partial biodegradability \cite{jimenez2012edible,attaran2017materials}. However, the different polarities of TPS and PE lead to phase separation, resulting in a material with poor mechanical and barrier properties \cite{Wang2015, Mao2012, altayan2022toward}. The compatibilization of polymer blends can be achieved by using nanoparticles \cite{Mao2012}, block copolymers \cite{Eastwood2002}, and phyllosilicate clays \cite{Huang2004, Rezaei2015}. Food contact approvals, high surface-to-volume ratio, and low cost of phyllosilicate clays, such as tetramethylammonium-montmorillonite (TMA-MMT), have made them an attractive option for compatibilization in food packaging applications \cite{Rezaei2015, zhou2015starch, Suter2015}. Additionally, \ce{Mg} substitutions in the octahedral or the tetrahedral layer of phyllosilicate clays create a permanent negative layer charge that facilitates cation exchange, enabling surface modification through replacement with either metallic (e.g., in K-MMT) or alkylammonium ions containing one or more C5-C18 alkyl units (referred to as organically modified MMT, oMMT). This surface modification has been shown to allow use of oMMT for compatibilization of various phase-separated polymer systems by promoting molecular interactions at the phase boundaries, thus reducing interfacial tension and leading to an improvement of mechanical properties such as the Young's modulus through improved stress transfer between phases \cite{Krishnamoorti1996, Cyras2008, Huang2004, Bher2019, Dutta2017, Rezaei2015}. For example, the addition of \SI{5}{\percent} cloisite-15A (an oMMT clay) in a TPS-PE blend decreased the diameter of TPS domains from \SI{500}{\nano\meter} to \SI{200}{\nano\meter}, resulting in a more uniform morphology and a \SI{35}{\percent} improvement in the tensile strength \cite{Sabetzadeh2016a,sharif2011study}. However, selecting an appropriate phyllosilicate clay for a specific polymer-clay nanocomposite (PCNC) system could be a time- and cost-intensive exercise, requiring optimization over a large parameter space of the polymer matrix composition and clay surface functionalization.\\
Molecular dynamics (MD) simulations allow accurate determination of the structure-property relationship of multicomponent systems, complementing material design by providing in silico screening over a large parameter space \cite{huang2022molecular, guo2023molecular, lin2023multiscale, gartner2019modeling}. MD simulation at an all-atom (AA) scale has shown its application in the accurate prediction of Young's modulus and glass transition temperature of TPS \cite{patidar2024martini, Ozeren2020}, the radius of gyration and density of PE \cite{panizon2015martini}, and mechanical properties of TMA-MMT clay \cite{Khan2019}. However, AA simulations become computationally prohibitive for most PCNCs systems of interest due to the combination of a small integration time step ($~ \SI{1}{\femto\second}$), long relaxation time of high molecular weight polymers, and at least one large dimension for highly anisotropic clay particles (diameter $\sim$\SI{200}{\nano\meter}). For example, AA simulation of a model TPS--TMA-MMT composite consisting of small polymer chains ($R_\mathrm{g}$$\sim$\SI{2}{\nano\meter}) and a small MMT sheet ($\SI{10.32}{\nano\meter} \times \SI{10.71}{\nano\meter} \times \SI{1.2}{\nano\meter}$) required $10^9$ steps to obtain convergence in polymer polymer near the clay surface \cite{patidar2024martini}.\\
In coarse-grained (CG) approaches, groups of atoms, monomers, or entire polymer chains are represented as a single bead/particle, significantly reducing the computational cost while capturing the essential properties relevant to the phenomenon under study. Some mesoscale methods like dissipative particle dynamics (DPD) simulations \cite{Groot1997,Singh2018,Fu2013,Scocchi2007,long2006nonlinear,sliozberg2020dissipative,khani2015polymer,wang2021dissipative,ju2013miscibility}, field-based theories \cite{schweizer1992reference, schweizer1997polymer, zirkel2002small, wu2006density,oxtoby2002density,helfand1975theory,helfand1976block,lowden1973solution} and integral-equation theories \cite{sung2005integral, hall2011impact,martin2016using, hsu1979rism,martin2018pyprism} have been used extensively to study morphological, structural and thermodynamics properties of extensive large systems ($10^3$\SI{}{\nano\meter}). These methods utilise a soft-core potential between larger chain units (Kuhn segments), which is determined from appropriate experimental data or AA simulations. However, a small chemical modification at the monomer level can completely alter the material morphology \cite{zhou2015starch}, and these changes cannot be directly captured by these methods. Additionally, the applicability of their parameters is limited to specific temperature and pressure ranges, and estimating them for multi-component formulations becomes increasingly challenging, making them inefficient models for the current system.\\

Chemically-specific CG models, like iterative Boltzmann inversion (IBI) \cite{reith2001mapping,reith2003deriving}, force matching \cite{izvekov2005multiscale} and MARTINI FF \cite{Marrink2007}, combine 4-10 atoms (bead size $\sim$\SI{}{\nano\meter}) in one bead, and therefore, provide better chemical representation with a $\sim 100$ $\times$ speed up over AA simulations. However, IBI and force matching methods do not offer good chemical transferability, and temperature transferability is obtained by optimizing the parameters using data from multiple state points \cite{moore2014derivation}. The MARTINI-FF, based on partitioning energy between polar and apolar solvents, provided good transferability across diverse chemical systems and molecular environments \cite{Marrink2007}, with proven application in estimating surface tension of solvents \cite{ndao2015coarse}, making it a suitable choice for interfacial studies. Previous studies have developed and validated the MARTINI-2 FF parameters for PE \cite{panizon2015martini}, TPS \cite{Lopez2009,patidar2024martini}, and phyllosilicate clay \cite{Khan2019}. The clay parameters successfully reproduced AA density profiles and chain-level properties of multiple polymers (polyethylene, polypropylene, and polystyrene) near TMA-MMT clay across various temperatures, demonstrating robust parameter transferability across diverse chemical environments and state points \cite{Khan2019}. A subsequent study further showed that clay interaction parameters derived from dispersive and polar contributions to clay sheet cleavage free energy at room temperature remain accurate at elevated temperatures, owing to a negligibly small entropic contribution to the cleavage free energy \cite{doi:10.1021/acs.jctc.5c01004} However, in biomolecular systems, the delicate enthalpy–entropy balance can lead to overstructuring and reduced thermodynamic transferability across different temperature conditions \cite{jarin2021coarse}. Additionally, one typical limitation of MARTINI-2 FF was a need for rescaling some of the cross-interactions to avoid freezing near the surface \cite{wu2012coarse,gobbo2013martini,piskorz2019nucleation, patidar2024martini}. The MARTINI-2 FF parameters of TPS components like sorbitol \cite{sukenik2015osmolyte} and polysaccharides \cite{Schmalhorst2017,shivgan2020extending} also required rescaling of the self-interaction term to improve agreement with AA simulations and experimental data. This underscores the critical need for systematic reparameterization of FF to establish a correct balance of solute-solvent interactions. The recently introduced MARTINI-3 FF parameters addressed these limitations, making it an ideal FF for studying the behavior of mixed or multi-component systems while avoiding the need for parameter rescaling \cite{souza2021martini, wang2025martini3, alessandri2022martini, thomasen2022improving}. MARTINI-3 models for several carbohydrates including linear 1--4 $\alpha$-D glucose and branched 1--6 $\alpha$-D glucose linkages are available \cite{lutsyk2022extending, grunewald2022polyply}. However, to the best of our knowledge, the MARTINI-3 model of pyrophyllite or any surface-modified MMT is not available.\\   
Given the wide interest and potential applications of phyllosilicate-based composites, we have developed MARTINI-3 CG FF parameters for pyrophyllite and TMA-MMT clay using a combination of experimental and AA MD simulation data. The hydration energies obtained from experimental surface tension data were used for initial surface bead assignments of pyrophyllite, which were fine-tuned based on optimal reproduction of AA structural (radius of gyration), thermodynamic (two-body excess entropy, density, conformational entropy), and dynamic properties (diffusion coefficient) of TPS near pyrophyllite clay. The same set of AA properties in TPS--[tetramethylammonium (TMA)]-MMT composite was used to determine the optimal bead assignments for the new functional groups of TMA-MMT resulting from Al to Mg substitution in the octahedral layer of pyrophyllite. The developed CG model was used to investigate the effect of dodecyltrimethylammonium (C12TAB)-modified montmorillonite (MMT), an organically-modified MMT with one C12 alkyl chain substitution on the nitrogen, on the morphology and interfacial properties of a large TPS-PE system. The morphological insights (clay assemblies in bulk and at the interface) and the effects of clay on the interfacial surface tension of the TPS-PE phase were compared with SEM images and XRD data for systems with similar compositions \cite{Sabetzadeh2016,sharif2011study}.

\section{Methods}
\subsection{System and Simulation Details}
AA and CG simulations of TPS melt (using the pre-existing CG model \cite{lutsyk2022extending}, with bead mapping shown in Figure \ref{fig:mapping}), TPS-pyrophyllite (TPS-PY), and TPS--TMA-MMT composites were employed to refine the initial bead assignments for pyrophyllite and TMA-MMT clay surfaces. The fine-tuned parameters were subsequently utilized for simulations of larger systems, including TPS-PE melt and TPS-PE-Clay (TPS-PE-CL) composites. Table \ref{tab:system details} provides more information on these systems.   

\begin{table*}[!htb]
  \centering
  \resizebox{\textwidth}{!}{%
 \setlength{\tabcolsep}{0.5pt}
  \captionof{table}{\textbf{System details.} Composition and simulation details for each studied system, including weight percentage (wt \SI{}{\percent}), number of molecules for each component, and total number of beads used in AA ($n_{AA}$) and CG ($n_{AA}$) simulations. All TPS composite systems were equilibrated via annealing cycles between \SI{613}{\kelvin} and \SI{913}{\kelvin} at multiple pressures (see Supplementary Information Section \ref{annealing} for more details), followed by NVT (\SI{25}{\nano\second}) and NPT (\SI{100}{\nano\second}) runs at \SI{613}{\kelvin}, with production MD simulations (1000-1500 \SI{}{\nano\second}) performed at \SI{613}{\kelvin} and \SI{1}{\bar}. Solvent systems (hexadecane, propanol, ethanol, isopropanol, and sorbitol) and their composites with periodic clay (pyrophyllite and TMA-MMT) were simulated sequentially through NVT (\SI{25}{\nano\second}), NPT (\SI{100}{\nano\second}), and production MD (\SI{1000}{\nano\second}) at \SI{300}{\kelvin} and \SI{1}{\bar}, except sorbitol-containing systems, which were simulated at \SI{450}{\kelvin} and \SI{1}{\bar}.} 
  \centering
  \label{tab:system details}
  \begin{tabular}{c c c c c c}
   \toprule
    System  & components (wt \SI{}{\percent}, Number of molecules) &   & $n_{AA}$ & & $n_{CG}$ \\
    \toprule   
    TPS  & amylopectin (\SI{50}{\percent}, 35), amylose (\SI{20}{\percent}, 28), sorbitol (\SI{29}{\percent}, 671), water (\SI{1}{\percent}, 54) &  & 55327 & & 9123 \\
    TPS-PY  & TPS (\SI{80}{\percent}, 1576), periodic pyrophyllite clay (\SI{20}{\percent}, 1) & & 120254 & & 21126 \\
    TPS--TMA-MMT  & TPS (\SI{82.5}{\percent}, 1576), periodic TMA-MMT clay (\SI{17.5}{\percent}, 1) &  & 122974 &  &21286 \\
    TPS-PE  & TPS without water (\SI{50}{\percent}, 3638) , PE (\SI{50}{\percent}, 108) & & - &&  338688 \\
    TPS-PE-CL  & TPS without water (\SI{47.5}{\percent}, 3638), PE (\SI{47.5}{\percent}, 108), C12TAB-MMT clay (\SI{5}{\percent}, 8) & & - & & 366848 \\
    PE-CL  & PE (\SI{97.5}{\percent}, 108), C12TAB-MMT clay (\SI{2.5}{\percent}, 2) & & - & & 169040 \\
    Solvent  &  & & 67650-102432 & & 5064-8928 \\
    Solvent-Pyrophyllite  & solvent, periodic pyrophyllite clay & & 67250-103650 & & 10106-13330 \\
    Solvent-TMA-MMT  & solvent, periodic TMA-MMT clay & & 78270-118601 &  &10240-15240 \\
    \bottomrule
 \end{tabular}}  
 \end{table*}
\subsubsection{AA System and Simulation Details}
\begin{figure*}[!htb]
     \centering
     \includegraphics[width=10cm,height=12cm]{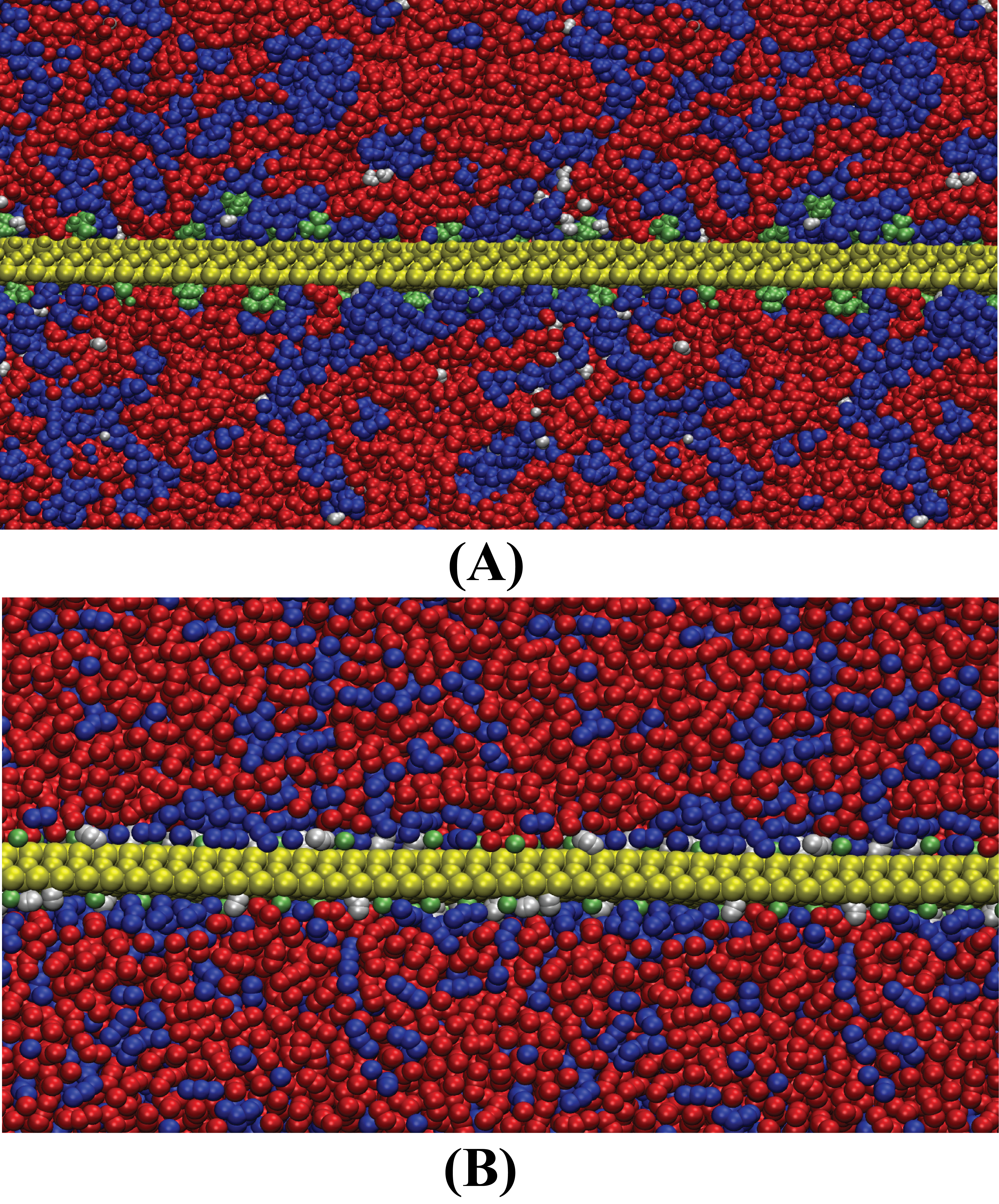}
     \caption{\textbf{AA and CG representation of TPS-TMA--MMT composite system}. \textbf{(a)} shows the AA representation of the TPS--TMA-MMT composite system. \textbf{(b)} shows the same system in CG representation. The polymer (red), sorbitol (blue), water (white), sheet (yellow), and TMA ion (lime) beads are represented by different colors.}
     \label{fig:AACGcomposite}
 \end{figure*}

AA structures and FF parameters of amylose (modeled as a linear chain with eighteen $\alpha$-D glucose monomers), amylopectin (modeled as a primary chain comprising twenty-seven $\alpha$-D glucose monomers and a secondary branch composed of nine monomers), and sorbitol were obtained using CHARMM-GUI \cite{jo2008charmm, guvench7ii, guvench2009charmm}. The INTERFACE FF was used to model pyrophyllite, and TMA-MMT \cite{heinz2005force}, and water was modeled as TIP3P \cite{sun1995hydrophobic}. For TMA-MMT, octahedral $\ce{Al^{3+}}$ were replaced isomorphically by $\ce{Mg^{2+}}$ without forming $\ce{Mg-(OH)_2-Mg}$ pairs, giving a unit cell formula of $\ce{[TMA]_{0.33}[Si_4O_8][Al_{1.66}7Mg_{0.333}O_2(OH)_2]}$ for a charge exchange capacity of \SI{0.9}{\milli\mol\per\gram}).\\
Packmol \cite{martinez2009packmol} was used to obtain a low-density TPS system ($\rho = \SI{117.62}{\kilo\gram\per\cubic\meter}$), which was subjected to multiple box rescaling and temperature-pressure ($T$-$P$) annealing cycles to obtain a high-density (\SI{1227.62}{\kilo\gram\per\cubic\meter}) system (more details provided in Supplementary Information section \ref{annealing}). The TPS composite systems consisting of pyrophyllite (TPS-PY) and TMA-MMT (TPS--TMA-MMT) sheets were obtained by sandwiching respective periodic sheets ($\SI{10.32}{\nano\meter} \times \SI{10.71}{\nano\meter} \times \SI{1.2}{\nano\meter}$ connected across $x$-$y$ periodic boundaries) between two pre-equilibrated TPS melt boxes (an equilibrated morphology TPS--TMA-MMT system is shown in Figure \ref{fig:AACGcomposite} (a)).\\
AA MD simulations were performed using GROMACS 2024.2 \cite{abraham2015gromacs}, with a time step of \SI{1}{\femto\second}, V-rescale thermostat ($\tau$=\SI{1}{\pico\second}), and a combination of the Berendsen barostat ($\tau$=\SI{2}{\pico\second}, first \SI{100}{\nano\second}) and the Parrinello-Rahman barostat ($\tau$=\SI{5}{\pico\second}).
\subsubsection{CG System and Simulation Details}
\begin{figure}[!htb]
     \centering
     \includegraphics[width=6.35cm,height=10.20cm]{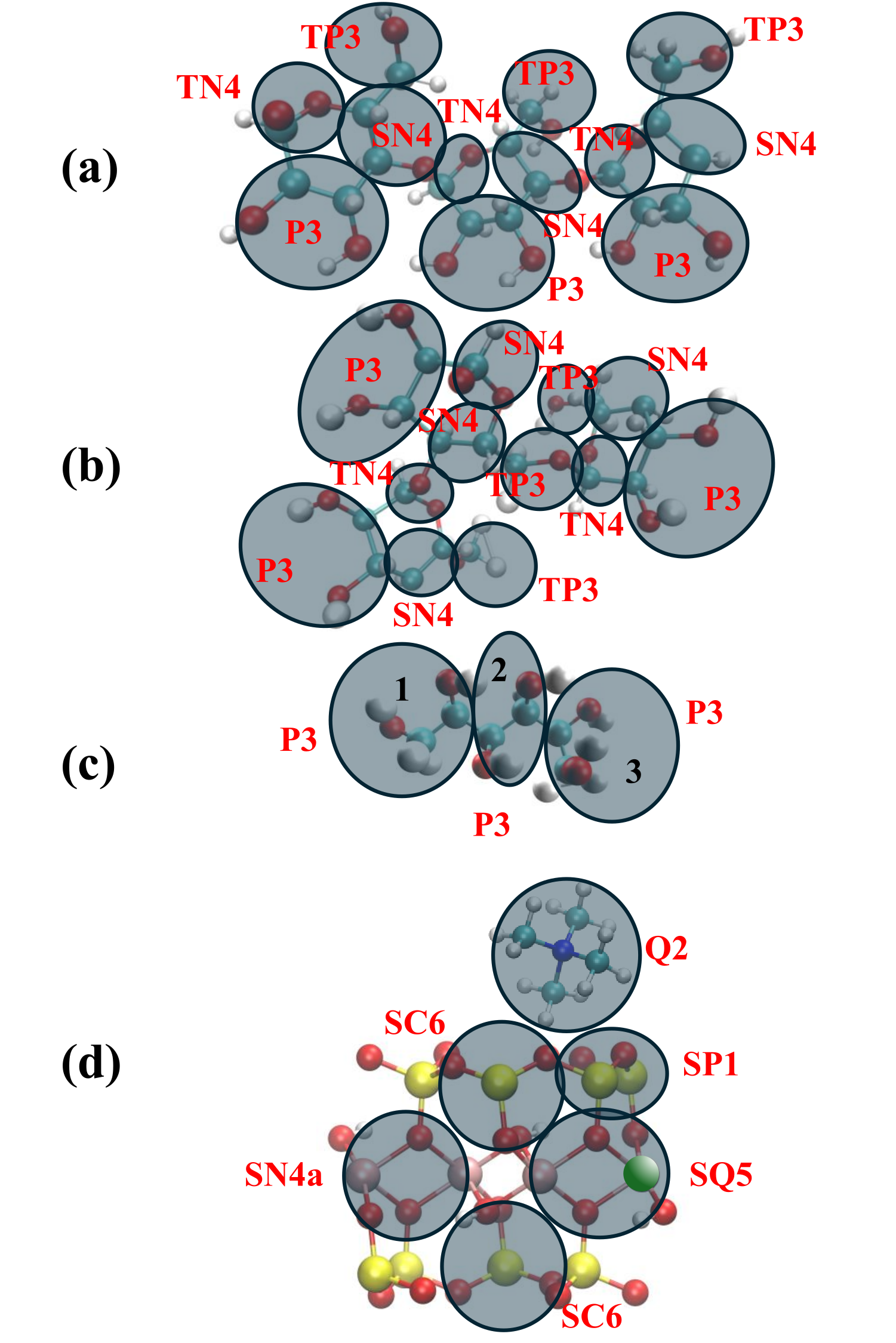}
     \caption{ \textbf{CG MARTINI-3 mapping scheme for TPS and TMA-MMT clay.} Atom groups and bead types for: \textbf{(a)} a trimeric segment of $\alpha$-D-glucose units connected with 1--4 linkage (representing a small unit of linear amylose chain), \textbf{(b)} 1--4 linked dimeric segment of $\alpha$-D-glucose with an additional 1--6 connection (represent branching in amylopectin), \textbf{(c)} sorbitol, and \textbf{(d)} a TMA-MMT unit cell with one octahedral $\ce{Al}$-for-$\ce{Mg}$ substitution, resulting in $-1 \, e$ charge balanced by $+1 \, e$ TMA ion. Pyrophyllite clay contains no such substitutions or counter-ions. Red, white, cyan, yellow, pink, and green represent oxygen, hydrogen, carbon, silicon, aluminum, and magnesium atoms, respectively. Atoms within shaded contours are represented by a single CG bead, with MARTINI-3 bead type specified. Chemical structure diagrams with explicitly labelled atoms are provided in Supplementary Information Figure \ref{fig:mappingchemical}, and complete atom-to-bead assignments are listed in Supplementary Information Table \ref{tab:mappingchemical}.}
     \label{fig:mapping}
 \end{figure}

\begin{figure*}[!htb]
     \centering
     \includegraphics[width=14cm,height=5.5 cm]{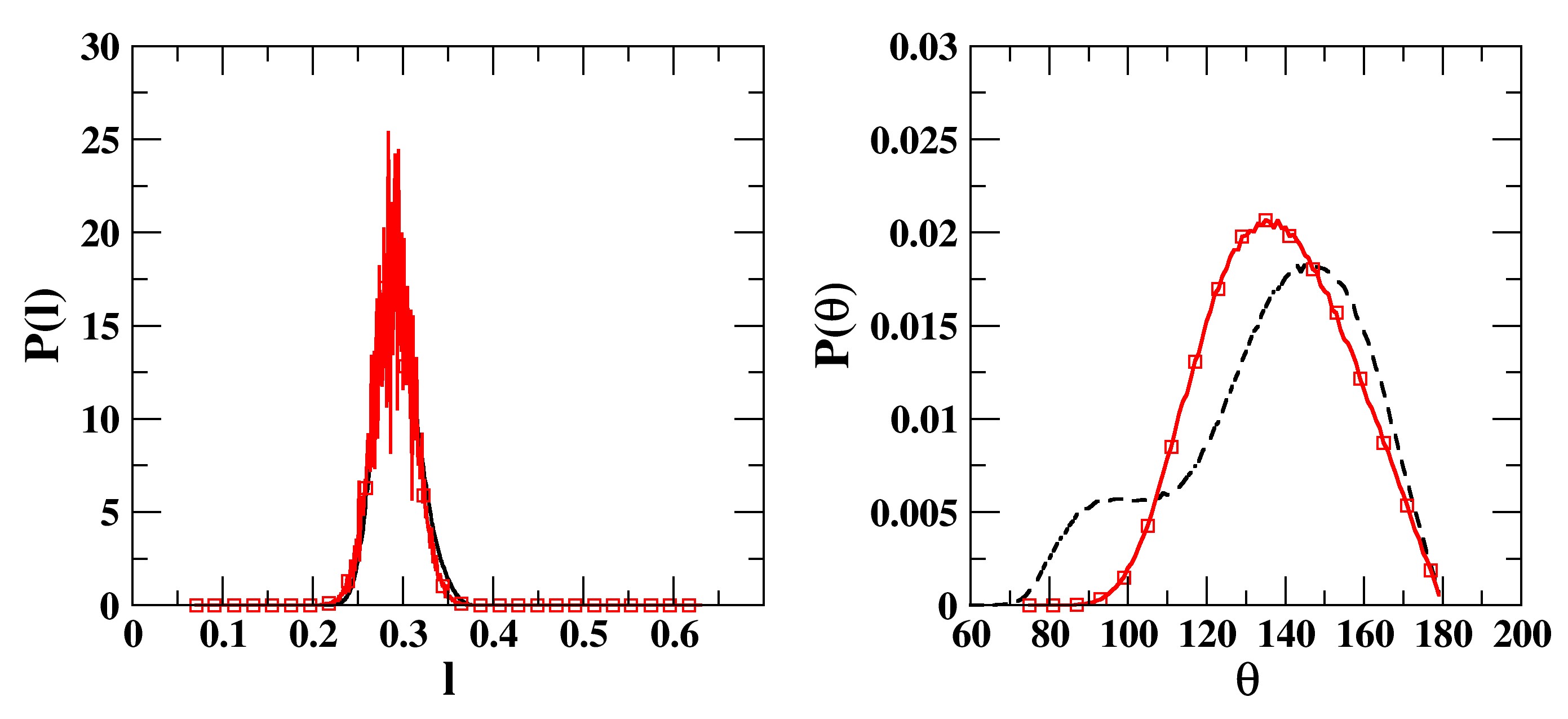}
     \caption{\textbf{Bonded distributions of sorbitol in AA and CG MARTINI-3 simulations}. \textbf{(a)} The bond, and \textbf{(b)} angle distribution of sorbitol, both obtained from the AA (\blackdotline) and CG MARTINI-3 (\redwithsquare) simulations of TPS melt at \SI{613}{\kelvin}.}
     \label{fig:bondsorb}
 \end{figure*}

\begin{figure*}[!htb]
     \centering
     \includegraphics[width=11cm,height=14 cm]{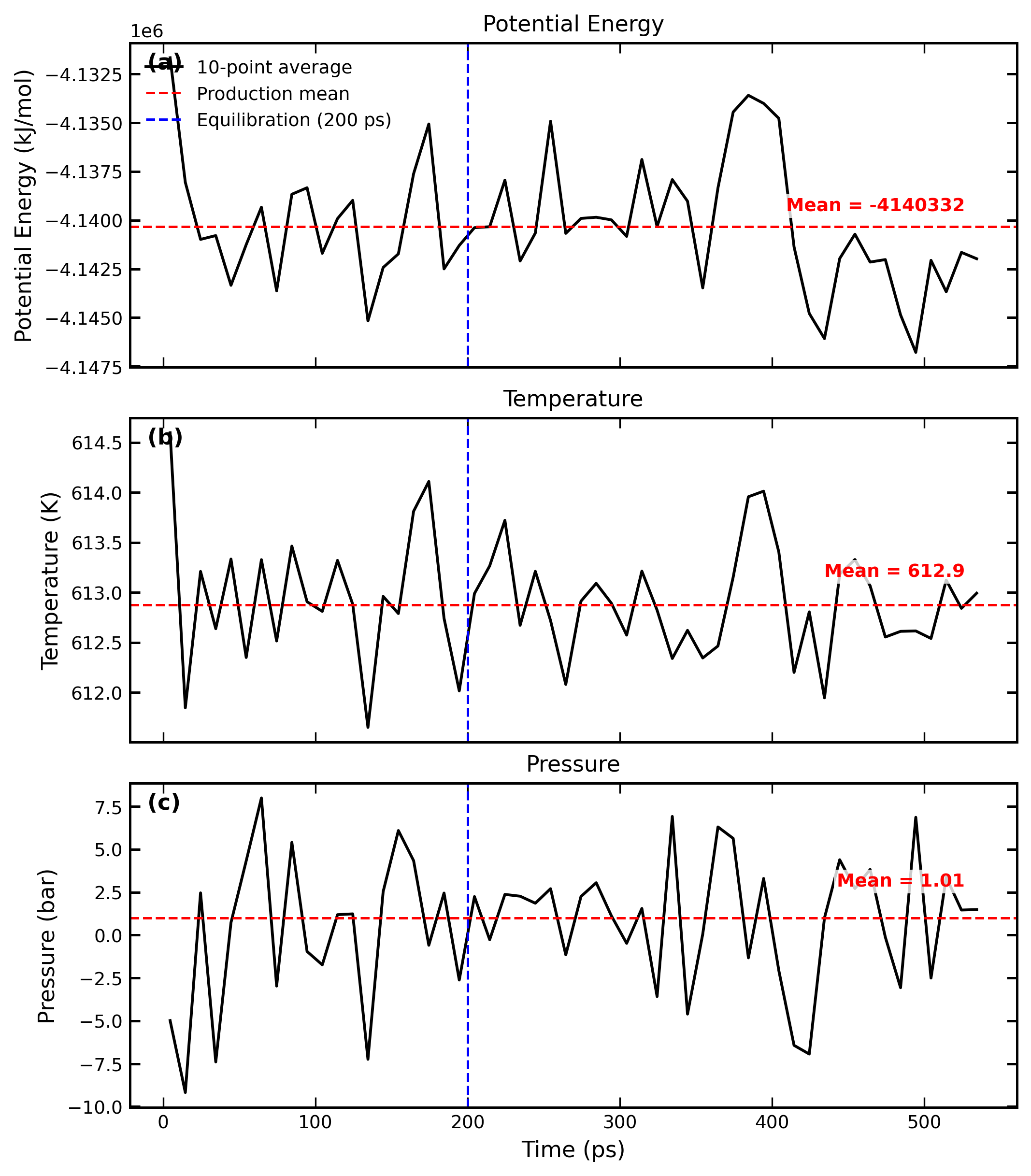}
     \caption{\textbf{Equilibration properties for a \SI{550}{\pico\second} CG MARTINI-3 simulation of TPS-PE-CL system using \SI{15}{\femto\second} time step.} \textbf{(a)} Potential energy, \textbf{(b)} temperature, and \textbf{(c)} pressure values of TPS-PE-CL system at \SI{613}{\kelvin} and \SI{1}{\bar}.}
     \label{fig:15fseq}
 \end{figure*}

\begin{figure*}[!htb]
     \centering
     \includegraphics[width=14cm,height=8cm]{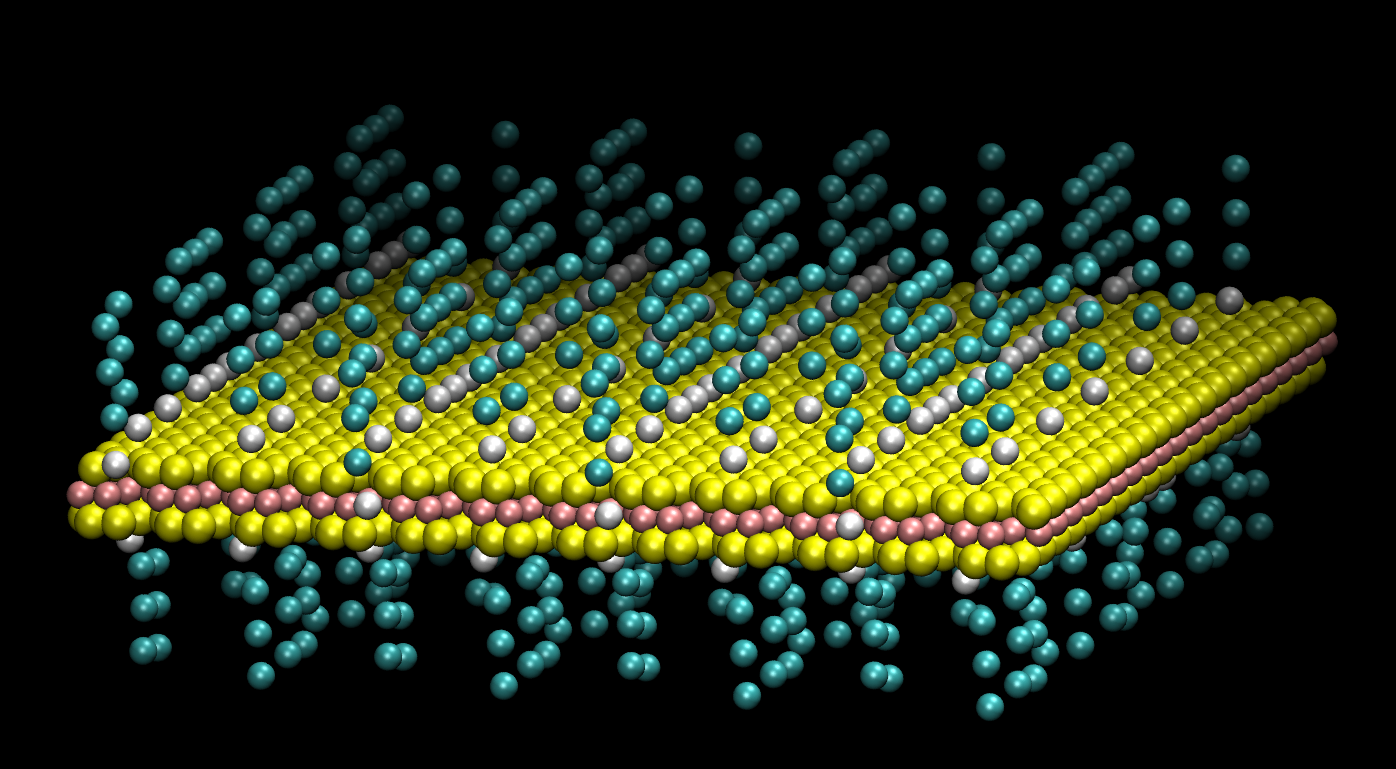}
     \caption{\textbf{Structure of C12TAB-MMT sheet} The C12 graft, depicted as three C1 type CG beads (cyan color), was grown on the TMA (white color beads) within a well-equilibrated TMA-MMT structure. The surface ($\ce{SiO_2}$) and internal beads ($\ce{AlO2H}$ and $\ce{MgO2H}$) of the sheet are represented by yellow and pink colors, respectively.} \label{fig:grafing}
 \end{figure*}

\begin{figure*}[!htb]
     \centering     \includegraphics[width=14cm,height=7cm]{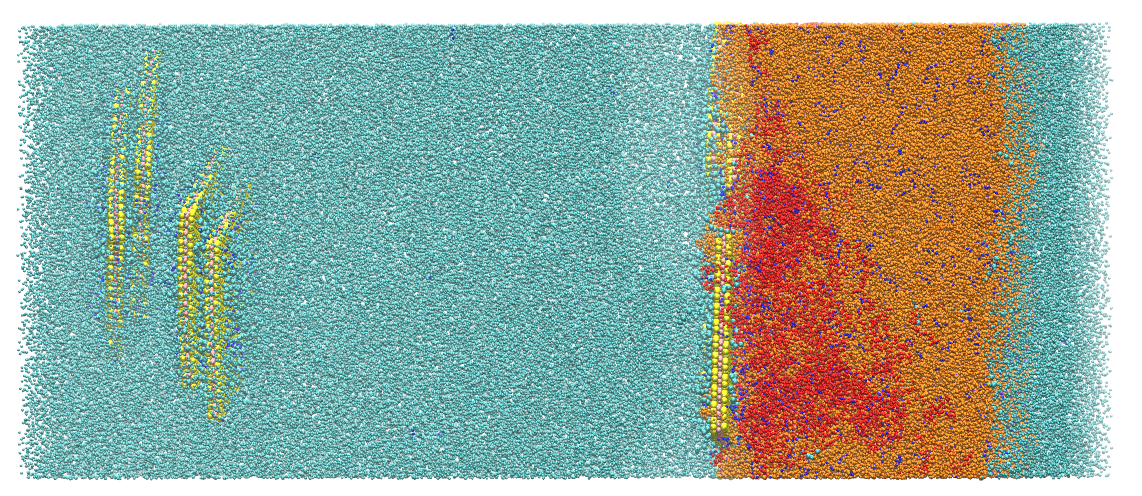}
     \caption{\textbf{TPS-PE-CL system} The equilibrated morphology of TPS-PE-CL system obtained at the end of temperature-pressure annealing cycles. The C12TAB-MMT surface $\mathrm{SiO_2}$, internal $\mathrm{MgO_2H}$ or $\mathrm{AlO_2H}$, and graft (TMA and C12 alkyl chains) CG beads are represented by yellow, pink, and cyan colors, respectively. The amylose, amylopectin, and sorbitol CG beads are represented by red, orange, and blue colors, respectively.} \label{fig:initialTPSPECL}
 \end{figure*}

The MARTINI CG potential is given as per Equation \ref{eq:LJ}, 
\begin{multline}\label{eq:LJ}
U=4\epsilon_{ij}\left[\left(\frac{\sigma_{ij}}{r_{ij}}\right)^{12}-\left(\frac{\sigma_{ij}}{r_{ij}}\right)^6\right]+\frac{q_{i}q_{j}}{4{\pi}\epsilon_0\epsilon_{rel}{r_{ij}}}
+\frac{1}{2}K_l(l_{ij}-l_0)^2+\frac{1}{2}K_\theta(cos\theta_{ijk}-cos\theta_0)^2\\+K_\phi[1+(cos\phi_{ijkl}-cos\phi_0)]
\end{multline}
where $\sigma_{ij}$, $\epsilon_{ij}$, $q_i$ and $q_j$, and $\epsilon_{rel}$ are bead diameter, interaction strength, bead charge, and relative dielectric constant, respectively. The force constants for bond ($l_{ij}$), angle ($\theta_{ijk}$), and torsion ($\phi_{ijkl}$) potentials are $K_l$, $K_\theta$, and $K_\phi$, respectively. Within MARTINI-3 FF, $\sigma_{ij}$ values can vary from 0.34 to 0.62 depending on the size of interacting beads, which could be regular, small, or tiny. The choice of $\epsilon_{ij}$ is linked to the functional group(s) in the CG bead, with primary categorization defined as polar (P), intermediate/nonpolar (N), apolar (C), halo-compounds (X), monovalent ions (Q), divalent ions (D), or water (W). Except for W and D beads, all beads have subtypes characterized by an integer value ranging from 1 to 6, which indicates their interaction strength.\\ 
The MARTINI FF parameters for amylose and amylopectin were taken from the MARTINI-3 polysaccharide model, parameterized using AA simulations of mono- and disaccharides in polar and apolar solvents (extensively tested in temperature range of \SI{300}{\kelvin} to \SI{350}{\kelvin}) \cite{lutsyk2022extending}. Accordingly, we used the following bead assignments: (tiny) $\ce{TP_3}$ bead for hydroxymethyl, (small) $\ce{SN_4}$ for the ring oxygen, anomeric carbon, and the reduced-end anomeric hydroxyl, $\ce{P_3}$ for diol, and $\ce{SN_4}$ for the hydroxyethyl groups (see Figure \ref{fig:mapping} for the mapping scheme). The applicability of these CG parameters at \SI{613}{\kelvin} were extensively tested by comparing the AA properties of TPS melt against AA simulation data.
Sorbitol was modeled using three $\ce{P_3}$ beads, each representing a diol group (Figure \ref{fig:mapping} (c)). The statistics for the center of geometry (COG) positions of these diol groups in AA simulations were used to estimate bond and angle parameters in the CG model of sorbitol (see Figure \ref{fig:bondsorb} for distribution). The CG structure and parameters of PE were generated using Polyply software \cite{grunewald2022polyply}. The CG bonded parameters and mapping scheme (see Figure \ref{fig:mapping}) for the clay particles were adopted from our previous work \cite{khan2025characterization}, showing satisfactory agreement between density functional theory calculations (DFT), and AA and CG simulation results for TMA-MMT Young's modulus and bending stiffness. The functional groups in the internal octahedral layer ($\ce{AlO_2H}$ and $\ce{MgO_2H}$) and at the edge ($\ce{AlOH}$, $\mathrm{AL(OH)_2}$, $\mathrm{MgOH_2}$, and $\mathrm{Mg(OH)_2}$) were assigned according to MARTINI-3 rules (more details in section \ref{clay non-bonded parameters}) \cite{souza2021martini}. The bead-type for the surface $\mathrm{SIO_2}$ of pyrophyllite was selected based on hydration energies, which were calculated from experimental contact angle values of polar and apolar liquids on the pyrophyllite surface. More details on parameterization are provided in Section \ref{pyrophyllite non-bonded parameters}, with surface hydration energy calculations described in Section \ref{surface hydration energies}. This MARTINI model of pyrophyllite was tested by comparing the structural arrangement and dynamic behavior of hexadecane, propanol, ethanol, isopropanol, and sorbitol near the surface to AA simulation results. CG simulations using various assignments for clay surface beads, viz. $\ce{SC_4}$, $\ce{SC_6}$, $\ce{SN_2}$, $\ce{SN_4}$, $\ce{SN_6}$, and $\ce{SP_2}$, were performed for each solvent-pyrophyllite systems. Additionally, TPS-pyrophyllite (TPS-PY) nanocomposite simulations were conducted using these surface beads of varying polarity to systematically compare structural, thermodynamic, and dynamic properties with those obtained from AA simulations. Furthermore, we utilised the same AA properties obtained from TPS--TMA-MMT composite simulations to refine the MMT surface CG bead assignments, thereby capturing the induced surface polarity resulting from the substitution of $\ce{Al^{3+}}$ with $\ce{Mg^{2+}}$ in the octahedral layer.\\
The CG systems of TPS melt, TPS-PY, and TPS--TMA-MMT were prepared by placing CG beads at the COG positions (using Equation \ref{eq:mapping}) of constituent atoms (mapping scheme in Figure \ref{fig:mapping}) in a corresponding equilibrated AA structure. The required number of CG water beads were then added from a pre-equilibrated CG water box using the 'solvate' tool in GROMACS. A 50-50 wt\SI{}{\percent} phase-separated TPS-PE melt CG system was prepared in Packmol using polymer chains with sizes above their respective entanglement lengths ($N\geq\ce{N_e}$): amylose ($N=300, \ce{M_W}=\SI{48,642}{\gram\per\mole}, \ce{N_e} \sim 182$) \cite{jauregui1993onset}, amylopectin ($N=1300, \ce{M_W}=\SI{194,658}{\gram\per\mole}, \ce{N_e} \sim 600$) \cite{carriere1998evaluation}, and polyethylene ($N=3000, \ce{M_W}=\SI{42,081}{\gram\per\mole}, \ce{N_e} \sim 400$) \cite{nafar2019individual,foteinopoulou2009structure} (see Table \ref{tab:system details} for details). This system contains a branched amylopectin structure where linear chains of sixteen and thirty-six $\alpha$-D-glucose monomers were grafted to the main chain at ten- and twenty-monomer intervals, respectively. The TPS composition in this system was the same as the TPS melt system (see Table \ref{tab:system details}), except that the water fraction was replaced with sorbitol. A $\SI{10.32}{\nano\meter}\times\SI{10.71}{\nano\meter}$ C12TAB-MMT clay was prepared by grafting three $\ce{C_1}$ beads (represent a C12-alkyl unit) on a $\ce{Q_2}$ bead (TMA ion \cite{souza2021martini}). The equilibrated structure of this sheet was used to form a two-clay tactoid assembly (Figure \ref{fig:grafing},), used subsequently for insertion in the polymer blend system. A TPS-PE-CL (5 phr) composite system was prepared by adding C12TAB-MMT clay tactoids to the prepared TPS-PE system, wherein, two tactoids were positioned at the TPS-PE interface and two within the PE bulk. This overall assembly is representative of an experimentally observed TPS-PE blend with a similar oMMT filler \cite{sharif2011study, sabetzadeh2016effect}. The TPS-PE melt and TPS-PE-CL composite systems obtained from Packmol were subject to volume rescaling and temperature-pressure (T-P) annealing cycles to obtain high-density systems of $\SI{24.11}{\nano\meter} \times \SI{24.59}{\nano\meter} \times \SI{53.12}{\nano\meter}$ and $\SI{24.82}{\nano\meter} \times \SI{24.58}{\nano\meter} \times \SI{57.15}{\nano\meter}$, respectively, ensuring appreciable chain diffusion and morphology evolution away from the initial structure (the equilibrated TPS-PE-CL composite system is shown in Figure \ref{fig:initialTPSPECL}). The system was annealed between \SI{613}{\kelvin} and \SI{913}{\kelvin} with steps of \SI{50}{\kelvin}, wherein at each step, the system was energy minimized and subject to a \SI{5}{\nano\second} NPT simulation for local equilibration. These temperature cycles were performed at various pressure values between \SI{1}{\bar} to \SI{1000}{\bar}, with a $\sim 2$x increase up to \SI{1000}{\bar} followed by a $\sim2$x decrease back to \SI{1}{\bar} (see section \ref{annealing} for additional details).\\
All CG simulations employed the V-rescale thermostat ($\tau$ = \SI{1}{\pico\second}) with a time step of \SI{5}{\femto\second}. Although, all results reported in this work correspond to a \SI{5}{\femto\second} timestep in CG simulations, a \SI{15}{\femto\second} time step yielded stable trajectories post-equilibration (Supplementary Information Figure \ref{fig:15fseq}) making the developed parameters directly integrable with standard MARTINI CG simulations. The TPS melt, TPS--TMA-MMT, and TPS-PY composite systems were isotropically coupled using the Berendsen barostat ($\tau$ = \SI{8}{\pico\second}) \cite{jang2004molecular} for first \SI{100}{\nano\second} and the Parrinello-Rahman (PR) barostat ($\tau$ = \SI{40}{\pico\second}) \cite{lutsyk2022extending} thereafter. We employed semi-isotropic ($NP_zT$) pressure coupling, using the Berendsen barostat, for TPS-PE melt and TPS-PE-CL composite systems to enable accurate estimation of interfacial properties. The Berendsen barostat does not generate a rigorously correct ensemble. However, it remains the standard choice for semi-isotropic systems and interfacial tension calculation in MARTINI literature \cite{ndao2015coarse}.
\subsection{Property Calculation}
In the TPS--TMA-MMT and TPS-PY composite systems, the near- and far-sheet regions were classified based on density profiles of polymer and sorbitol along the normal ($z$-axis) to the surface (marked by the dashed red vertical line in Figure \ref{fig:de}). The COG distance of the polymer's trimeric segments ($\sim$ persistence length (\SI{0.86}{\nano\meter}) of amylose) from the sheet was used to classify it as a near or a far segment, with a chain labeled as a near/far-region chain if $>\SI{70}{\percent}$ segments were in the near/far region. All properties of the melts and composites were evaluated at \SI{613}{\kelvin}, while the solvent properties were analyzed at \SI{300}{\kelvin} (sorbitol properties were analyzed at \SI{450}{\kelvin}).

\begin{figure*}[!htb]
     \centering
     \includegraphics[width=17cm,height=7cm]{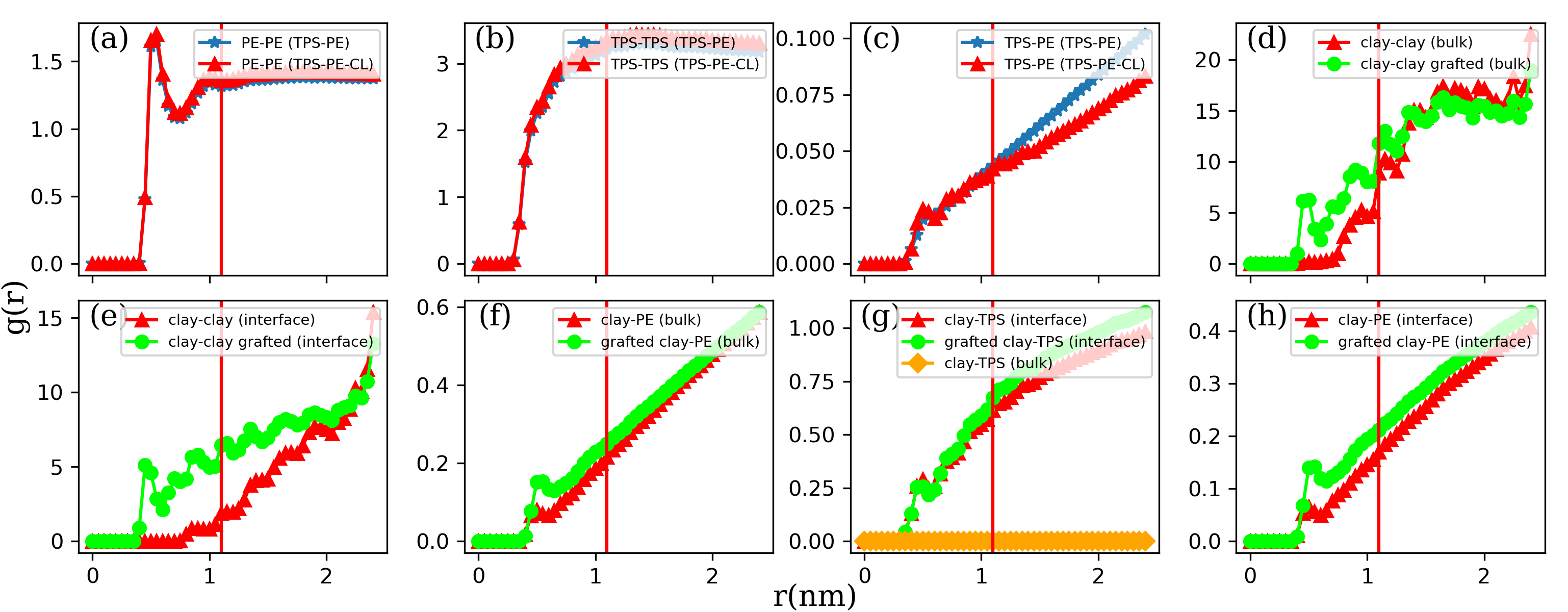}
     \caption{ \textbf{RDF of all possible pairs among TPS, PE, and clay in TPS-PE melt and TPS-PE-CL composite system}. Radial distribution functions (RDFs) obtained from coarse-grained (CG) MARTINI simulations of TPS-PE melt (\bluestar) and TPS-PE-CL composite (\redtriiangle, \greencircle, and \orangediamond) at \SI{613}{\kelvin} and \SI{1}{\bar}. Panels present RDFs for: \textbf{(a)} PE-PE, \textbf{(b)} TPS-TPS, \textbf{(c)} TPS-PE, \textbf{(d)} clay-clay (in PE bulk phase), \textbf{(e)} interfacial clay-clay, \textbf{(f)} clay-PE (in PE bulk phase), \textbf{(g)} interfacial clay-TPS and bulk clay-interfacial TPS (\orangediamond), \textbf{(h)} clay-PE (at TPS-PE interface), and \textbf{(i)} interfacial clay-PE pairs.}
     \label{fig:RDFTPSPECL}
 \end{figure*}

\paragraph{Structural and Dynamic Properties.}
We used in-built GROMACS utilities to calculate the density profiles ($\rho_n$), chain radius of gyration (${R_g}$), and radial distribution function (RDF) of polymer-sorbitol pairs in melt and composite systems. The density profile integrals, $Ar$ and $Ln$, were calculated using Equations \ref{eq:Areq} and \ref{eq:Lneq}, respectively, with $Ln$ providing the average distance of a species along the normal to the clay surface.
\begin{equation}\label{eq:Areq}
Ar = \int_{0}^{z} \rho(z)dz
\end{equation}
\begin{equation}\label{eq:Lneq}
Ln = \frac{1}{Ar}\int_{0}^{z} \rho(z)zdz
\end{equation}
The COG of atoms in a CG bead was used for RDF calculation in the AA simulations for comparison with the CG simulations. For the composite system, 2D RDFs were calculated in \SI{0.37}{\nano\meter} slices parallel to the TMA-MMT sheet. The RDFs are reported as averages over four slices in the near region and five slices in the far region. The coordination number (CN) of all possible pairs among TPS, PE, and CL was calculated as per Equation \ref{eqn:CN}, 
\begin{equation}
    \mathrm{CN} = 4 \pi\rho_\beta \int_{0}^{r_{cut}} r^2 g_{\alpha\beta}(r) dr
  \label{eqn:CN}
\end{equation}
where, $g_{\alpha, \beta} (r)$ is the RDF of these individual pairs, $r_{cut}$ is the cut-off distance, and $\rho_\beta$ is the number density of the $\beta$ component. We used $r_{cut}=\SI{1.1}{\nano\meter}$, corresponding to the second minima in RDF (Figure \ref{fig:RDFTPSPECL}). The contributions from bonded atoms were removed by excluding five, three, and nine neighbors for TPS, PE, and CL, respectively. The CN data were obtained from three trajectory segments: small \SI{10}{\nano\second} portion from each annealing cycle (at \SI{613}{\kelvin}), full \SI{25}{\nano\second} $NVT$ trajectory, and \SI{50}{\nano\second} blocks spanning the entire \SI{1500}{\nano\second} simulation. 
Slow evolution of polymer chain level properties necessitated a larger block of \SIrange{100}{300}{\nano\second} for $S_\mathrm{c}$ and $R_\mathrm{g}$ calculations. The diffusion coefficients for trimeric segments of amylose and amylopectin, and the solvent were calculated using mean-square displacement (MSD(t)) data in accordance with Einstein's Equations \ref{eq:MSD} and \ref{eq:MSD2}. The normalized diffusion coefficients ($D_\mathrm{n}$) were determined as the ratio of lateral diffusion in the $x$-$y$ plane in the composite to that in the melt, $D_\mathrm{n} = D_\mathrm{composite}/D_\mathrm{melt}$. In this definition, $D_\mathrm{n}$ captures the effect of clay particle on the self-diffusion coefficient of a species, with normalization enabling comparison of the effect of clay particle on dynamics in AA and CG systems. 
\begin{equation}\label{eq:MSD}
\mathrm{MSD (t)} = \mathrm{6 \cdot D_{melt} \cdot t}
\end{equation}
\begin{equation}\label{eq:MSD2}
\mathrm{MSD (t)} = \mathrm{4 \cdot D_{composite} \cdot t}
\end{equation}
\paragraph{Thermodynamic Properties.}
The pairwise ($S_{2,\alpha\beta}$) and total ($S_\mathrm{2}$) two-body excess entropies were calculated using the corresponding RDFs ($g_{\alpha,\beta}$) as per Equations \ref{eqn:tbe} and \ref{eq:overalltbe} (integral calculated up to \SI{2.5}{\nano\meter}), respectively \cite{sharma2008estimating}.
\begin{multline}
   \mathrm{S_{2,\alpha\beta}}=-2\pi\rho\int_{0}^{\infty}g_{\alpha\beta}(r)\mathrm{ln}\,g_{\alpha\beta} (r)-
   [g_{\alpha\beta}(r)-1]r^2dr
  \label{eqn:tbe}
\end{multline}
\begin{equation}\label{eq:overalltbe}
\mathrm{S_{2}}=\sum_{\alpha,\beta}{x_\alpha x_\beta}\mathrm{S_{\alpha\beta}},
\end{equation}
where, $x_\alpha$ and $x_\beta$ are the mole fractions of the $\alpha$ and $\beta$ components, respectively. The $S_{2,\alpha\beta}$ of composite systems were estimated using Equation \ref{eqn:tbecomp}. The configurational entropy ($S_\mathrm{c}$) of amylose and amylopectin chains was calculated from the covariance matrix of each chain (after removing rigid-body translational and rotational motion) using Schlitter's equation \cite{schlitter1993estimation}, implemented with GROMACS utility anaeig.\\
\paragraph{Potential Mean Force Calculation}
To quantify the thermodynamic stability of clay tactoid assemblies, we computed the PMF for separation of two C12TAB-MMT clay sheets in PE using \SI{10}{\nano\second} umbrella sampling (US) simulations at \SI{613}{\kelvin}, with clay separation, $\xi$. Each clay sheet was partitioned into four spatial quadrants of approximately equal bead count ($\simeq$720 beads) based on the x-y positions of the bare clay beads, and harmonic restraints of \SI{1000}{\kilo\joule\per\mol\per\nano\meter\squared} (see Equation \ref{eq:umbrella}) were applied to center of the mass of these quadrant beads to maintain the target sheet separation in each window. Starting configurations for 38 umbrella sampling windows spanning $\xi = 1.3$--\SI{5.0}{\nano\meter} (\SI{0.1}{\nano\meter} intervals) were extracted from \SI{18.5}{\nano\meter} of a \SI{20}{\nano\meter} steered MD simulation, in which a two-clay tactoid (initial $\xi$ = \SI{1.30}{\nano\meter}) was pulled along $\xi$ at \SI{0.0002}{\nano\meter\per\pico\second}. PMF was determined using WHAM (last \SI{5}{\nano\second} US simulations) reconstruction and standard error was estimated using 200 bootstrap iterations. The cohesive free energy ($\Delta$ G) was obtained from the PMF plateau value of full sheet separation (\SI{4.8}{\nano\meter}), and solvent-accessible surface area of the bare pyrophyllite platelet (2A = \SI{225.98}{\nano\meter\squared}) were used to calculate the work of adhesion ($W=\frac{\Delta G}{2 A}$).

\paragraph{Interfacial Tension Calculations.}
The interfacial tensions of the TPS-PE melt and TPS-PE-CL composite were calculated using the method of Irving and Kirkwood (IK), as per Equation \ref{eqn:IK}) \cite{kirkwood1949statistical},
\begin{equation}
       \mathrm{\gamma}=\int_{-L_z/2}^{L_z/2}(P_{N}(z)-P_{T}(z))dz=\frac{1}{2}\langle P_N - P_T \rangle L_z
  \label{eqn:IK}
\end{equation}
where, $P_N(z)$=$P_{zz}(z)$ and $P_T(z)$=$\frac{1}{2}$($P_{xx}(z)$+$P_{yy}(z)$) are the normal and tangential pressure tensor components in a box of height $L_z$, respectively. The $\gamma$ values were calculated in blocks of \SI{100}{\nano\second} over a \SI{1500}{\nano\second} trajectory, with standard deviation calculated from the converged portion of each trajectory. Furthermore, the solvent-accessible surface area (SASA) of pyrophyllite clay in TPS (TPS-PY system), with the clay surface as the reference group and the TPS as the selection group, was evaluated over the course of the simulation using the built-in GROMACS utility \texttt{gmx sasa}.
\section{Results and Discussion}

\subsection{Physical Insights from AA Simulation of Composites}\label{AA results pyrophyllite and MMT composite}

\begin{figure}[!htb]
   \centering
     \includegraphics[width=7.5cm,height=8.20cm]{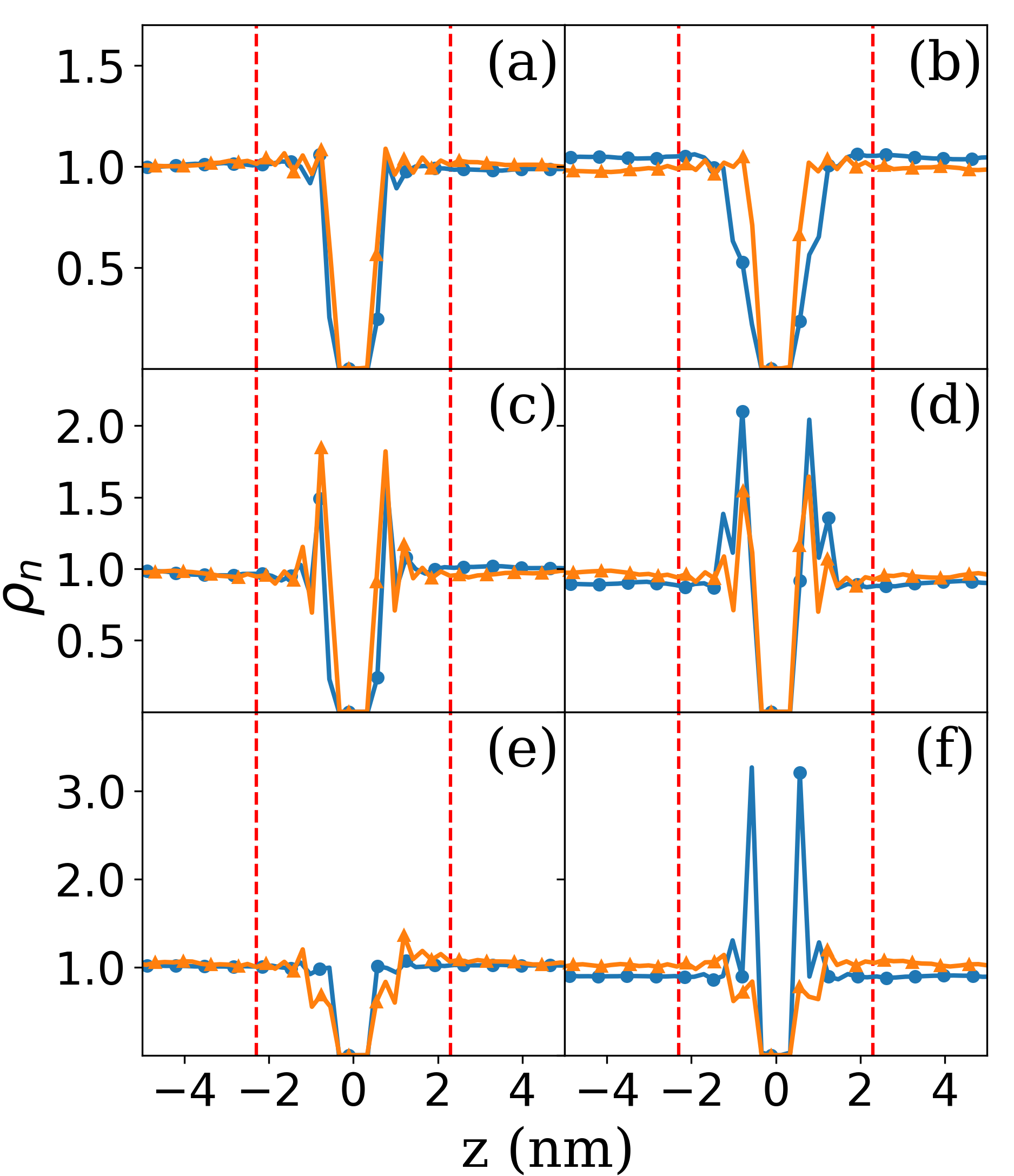}
     \caption{ \textbf{Normalized density profile of TPS-PY and TPS-MMT composite}. The density profiles of TPS components (polymer (amylose and amylopectin), sorbitol, and water) along the normal z-direction of the pyrophyllite \textbf{(a, c, e)} and TMA-MMT \textbf{(b, d, f)} sheets were obtained using AA (\bluewithcircle) and CG (\orangetriiangle) simulations of TPS composite systems. The obtained density profiles were normalized relative to the TPS melt density, with all simulations performed at $T$ = \SI{613}{\kelvin} and $P$ = \SI{1}{\bar}. The \reddotline indicates the boundary between the near-surface and far-surface regions. More details of systems are provided in Table \ref{tab:system details}.}
     \label{fig:de}
\end{figure}

Replacing the lower-polarity pyrophyllite surface with a higher-polarity TMA-MMT surface increased the extent of near-surface layering in the water and sorbitol distributions, accentuating starch polymer displacement in the first clay solvation layer (Figure \ref{fig:de}). This can be attributed to the preferential interaction of polar molecules (sorbitol and water) with charged TMA ions. A \SI{16.2}{\degree} decrease in water contact angle from pyrophyllite (\SI{79.2}{\degree}) to hectorite (structurally similar to TMA-MMT) aligns with this surface polarity increase \cite{giese1991surface}. The polymer and sorbitol components show a sharp increase ($\sim$ \SI{50}{\percent} of the bulk value) in density within the first solvation layer, driven by the interaction of water and sorbitol with the hydroxyl groups of amylose and amylopectin in the near-surface region, supported by experimental observation of increased clay swelling and TPS intercalation in the clay intergallery space with an increase in water content in a TPS–bentonite clay composite \cite{park2002preparation}.\\
The density profiles of all TPS components approach bulk values at an average $z$-distance of \SI{2.5}{\nano\meter} along the clay sheet normal, indicating negligible clay influence on TPS beyond this plane. Therefore, this plane was used to separate near and far clay regions for the subsequent calculation of all reported properties. The chain conformation was characterized using the radius of gyration, topological correlations using conformational entropy, and overall polymer dynamics using trimeric segment diffusion coefficients, all expected to show a significant dependence on polymer-surface interactions. These properties in the near-sheet regions, when normalized with respect to their corresponding values in the TPS melt, provided direct measures of clay influence on polymer behavior. Additionally, this normalization of diffusion coefficients enabled direct AA-CG comparison by removing the effects of inherently faster CG dynamics (w.r.t. AA) in both the melt and the composite systems resulting from a smoother free energy landscape for CG FF. The normalized diffusion coefficients ($D_\mathrm{n}$) for trimeric segments decreased from 1.30 (amylose) and 1.60 (amylopectin) near pyrophyllite to 0.96 and 1.16 near TMA-MMT, respectively (Tables \ref{tab:pyrophylliteRgScDn} and \ref{tab:TMSMMTRgScDn}). The near-to-melt $S_\mathrm{c}$ ratios showed a similar trend, decreasing from 1.03 (amylose) and 1.04 (amylopectin) near pyrophyllite to 0.99 and 0.92 near TMA-MMT, respectively. In contrast, near-to-melt $R_\mathrm{g}$ ratios were essentially the same for two surfaces. Generally, favorable polymer-surface interactions manifest as restricted segmental mobility (lower $D_\mathrm{n}$), reduced conformational freedom (lower $S_\mathrm{c}$), and chain expansion (increased $R_\mathrm{g}$), in near-surface regions (w.r.t. melt), whereas unfavorable interactions exhibit the opposite trends. For example, strong polymer-surface interactions caused a twofold $R_\mathrm{g}$ increase and reduced polymer mobility in polyethylene oxide/Na-MMT systems \cite{suter2009computer}. Weakly attractive interactions for polyethylene, polypropylene, and polystyrene near the TMA-MMT sheet lead to a \SI{10}{\percent} decrease in both $S_\mathrm{c}$ and diffusion coefficients \cite{Khan2019}. Conversely, repulsive nanoparticles accelerated Lennard-Jones (LJ) polymer dynamics \cite{smith2002molecular, varnik2002reduction} while causing negligible changes in $R_\mathrm{g}$ \cite{karatrantos2015polymer,smith2002molecular}. These observations indicate that pyrophyllite acts as a repulsive surface for starch polymers, while TMA-MMT acts close to a neutral surface. We conjecture that an effectively high plasticizer (sorbitol) concentration in the first solvation layer of TMA-MMT tempered the effect of slightly attractive TMA-MMT--TPS interactions, leading to only a small decrease in diffusion coefficients and conformational entropy of the polymer chains in the near TMA-MMT region w.r.t. to the melt values.\\

The AA simulations of these composites provided nanoscale insights into interfacial interactions and their relationship to the observed properties, but only for relatively simple systems, consisting of single periodic clay sheets and low $\ce{M_W}$ polymer chains (below the entanglement molecular weight), designed for faster equilibration. Modeling of more realistic polymer blend systems, such as starch-polyethylene composites studied by Korol et al.\cite{korol2015manufacture}, present a significantly greater complexity requiring extensive sampling of polymer conformational space, its packing near the clay surface, and clay reorganization at the interface, leading to prohibitive time- and length scales for AA simulations. This necessitates the development of chemically-specific CG parameters to access relevant scales while probing the role of specific molecular-level effects such as clay surface polarity.

\subsection{CG Parameter Development}

\begin{table*}[!htb]
 \small
  \centering
  \captionof{table}{\textbf{Comparison of properties obtained from AA and CG of TPS melt and TPS-PY composite system.} The TPS-PY CG simulations were run with surface $\mathrm{SiO_2}$ groups assigned various bead types in range $\ce{SC_4}$-$\ce{SP_2}$. The average values of radius of gyration ($R_\mathrm{g}$), configuration entropy per CG bead ($S_\mathrm{c}$), and normalized diffusion coefficient ($D_\mathrm{n}$) were calculated using the last \SI{900}{\nano\second} trajectories at \SI{613}{\kelvin} and \SI{1}{\bar}.}
  \centering
  \label{tab:pyrophylliteRgScDn}
\begin{threeparttable}
\resizebox{0.99\textwidth}{!}{
  \begin{tabular}{ c c c c c c c c c c } 
    \toprule
    
      \multicolumn{3}{c}{} & \multicolumn{4}{c} {\textbf{Amylose}} & \multicolumn{3}{c}{\textbf{Amylopectin}}   \\
    \cmidrule(lr){4-6}\cmidrule(lr){7-10}
      \multicolumn{2}{c}{} &\multicolumn{1}{c}{\textbf{bead-type}} &\multicolumn{1}{c}{\textbf{near}}  & \multicolumn{1}{c}{\textbf{far}}  &  \multicolumn{1}{c}{\textbf{melt}}  &   \multicolumn{1}{c}{} & \multicolumn{1}{c}{\textbf{near}} & \multicolumn{1}{c}{\textbf{far}} & \multicolumn{1}{c}{\textbf{melt}} \\
    \midrule
    $R_\mathrm{g}$(\SI{}{\nano\meter}) &    AA & & 1.34$\pm$0.07    & 1.35$\pm$0.08   & 1.38$\pm$0.02  &  & 1.72$\pm$0.11 & 1.80$\pm$0.10 & 1.83$\pm$0.02\\
     & CG      &  &    &    & 1.65$\pm$0.05 &  &  &  & 2.16$\pm$0.07 \\
    &       & $\ce{SC_4}$ &1.60$\pm$0.14    & 1.65$\pm$0.12   &  &  & 2.09$\pm$0.18 & 2.14$\pm$0.13 &  \\
    &       & $\ce{SC_6}$ & 1.59$\pm$0.13	&1.64$\pm$0.11 &	 && 2.07$\pm$0.20	& 2.15$\pm$0.15&	\\

    &      & $\ce{SN_2}$ & 1.62$\pm$0.11 & 1.62$\pm$0.09&	 && 2.12$\pm$0.16 & 2.16$\pm$0.13	&\\
    &       & $\ce{SN_4}$ & 1.57$\pm$0.12 & 1.64$\pm$0.10&	 && 2.07$\pm$0.18&	2.14$\pm$0.14	& \\
    &       & $\ce{SN_6}$ & 1.59$\pm$0.10	& 1.62$\pm$0.09&	 && 2.04$\pm$0.16	&2.15$\pm$0.16	&\\

    &       & $\ce{SP_2}$ & 1.57$\pm$0.13	& 1.63$\pm$0.09&	 && 2.05$\pm$0.15	&2.16$\pm$0.14	&\\

     \cmidrule(lr){3-10}
${\langle S_c \rangle}$  & AA  && 39.24$\pm$1.30  & 40.22 $\pm$1.16 & 37.97 $\pm$0.24 &  & 40.86$\pm$1.73 & 43.12$\pm$1.78 & 39.30 $\pm$0.29\\
$(\SI{}{\joule\per\mol\per\kelvin})$   & CG &      &     &    & 34.91 $\pm$0.22 &  &  &  & 36.87$\pm$0.19 \\
&& $\ce{SC_4}$     & 33.77 $\pm$1.40   & 36.34$\pm$1.29   &   &  & 37.10$\pm$1.65 & 38.29$\pm$1.70 &  \\
&&$\ce{SC_6}$&34.31$\pm$1.25&	36.18$\pm$1.17&	&& 36.23$\pm$1.80	&38.34$\pm$1.55	&\\
&&$\ce{SN_2}$&33.32$\pm$1.23&	35.72$\pm$1.18&	 && 36.50$\pm$1.50	& 39.00	$\pm$1.64& 
\\
&&$\ce{SN_4}$&33.66	$\pm$1.18&36.29$\pm$1.07&	&& 35.15$\pm$1.81&	39.28$\pm$1.74&	
\\
&&$\ce{SN_6}$&33.11$\pm$1.11&	36.46$\pm$1.26&	 && 34.66$\pm$1.60	& 37.82$\pm$1.52	& 
\\
&&$\ce{SP_2}$&34.01$\pm$1.29&	48.54$\pm$1.09& && 36.43$\pm$1.66&	39.25$\pm$1.49	& 
\\
\midrule

$D_\mathrm{n}$  & AA  && 1.30$\pm$ 0.21  & - & -  &  & 1.60$\pm$ 0.24 & - & - \\ 
& CG & $\ce{SC_4}$     & 0.75$\pm$ 0.10    & -   & -  &  & 0.84$\pm$ 0.11 & - & -\\
&&$\ce{SC_6}$& 0.74$\pm$ 0.09&	-&-&& 0.82$\pm$ 0.13	& -	& -\\
&&$\ce{SN_2}$&0.70$\pm$ 0.12&	-&	-&& 0.65$\pm$ 0.17	& -	& -\\
&&$\ce{SN_4}$& 0.23$\pm$ 0.18	& -&-&& 0.32$\pm$ 0.22&	-&	-\\
&&$\ce{SN_6}$&freezing&-	& -&& freezing	& -	& -\\
&&$\ce{SP_2}$&freezing&-	& -&& freezing	& -	& -\\
 \bottomrule
  \end{tabular}}
  \end{threeparttable}
\end{table*}

The properties obtained from AA simulations were used to refine the initial CG bead assignments for the clay surface, which were based on hydration energies calculated by applying the Lifshitz theory to experimental surface tension data (more details in Sections \ref{pyrophyllite non-bonded parameters} and \ref{surface hydration energies}). Simultaneous incorporation of multiple thermodynamic, structural, and interfacial properties in parameterization is expected to enhance the model's accuracy in capturing the complex free energy landscape of surface adsorption. For instance, the recently released MARTINI-3 FF demonstrated substantial improvements in solute-solvent interactions over earlier versions by integrating both partition-free energies and miscibility trend data, resulting in more transferable and reliable parameters \cite{souza2021martini,Marrink2007}. Following this approach, we validated our CG model through a hierarchical testing protocol. We first tested the CG model for the TPS system using AA simulation benchmarks, which established the baseline accuracy for polymer-polymer and polymer-sorbitol interactions. This was followed by progressively more complex TPS-PY and TPS-TMA-MMT composite simulations, which enabled surface bead refinement through direct comparison with AA properties.

\subsubsection{TPS Melt Properties from CG and AA simulations}\label{melt non-bonded paramters}
\begin{figure*}[!htb]
     \centering     \includegraphics[width=10cm,height=8cm]{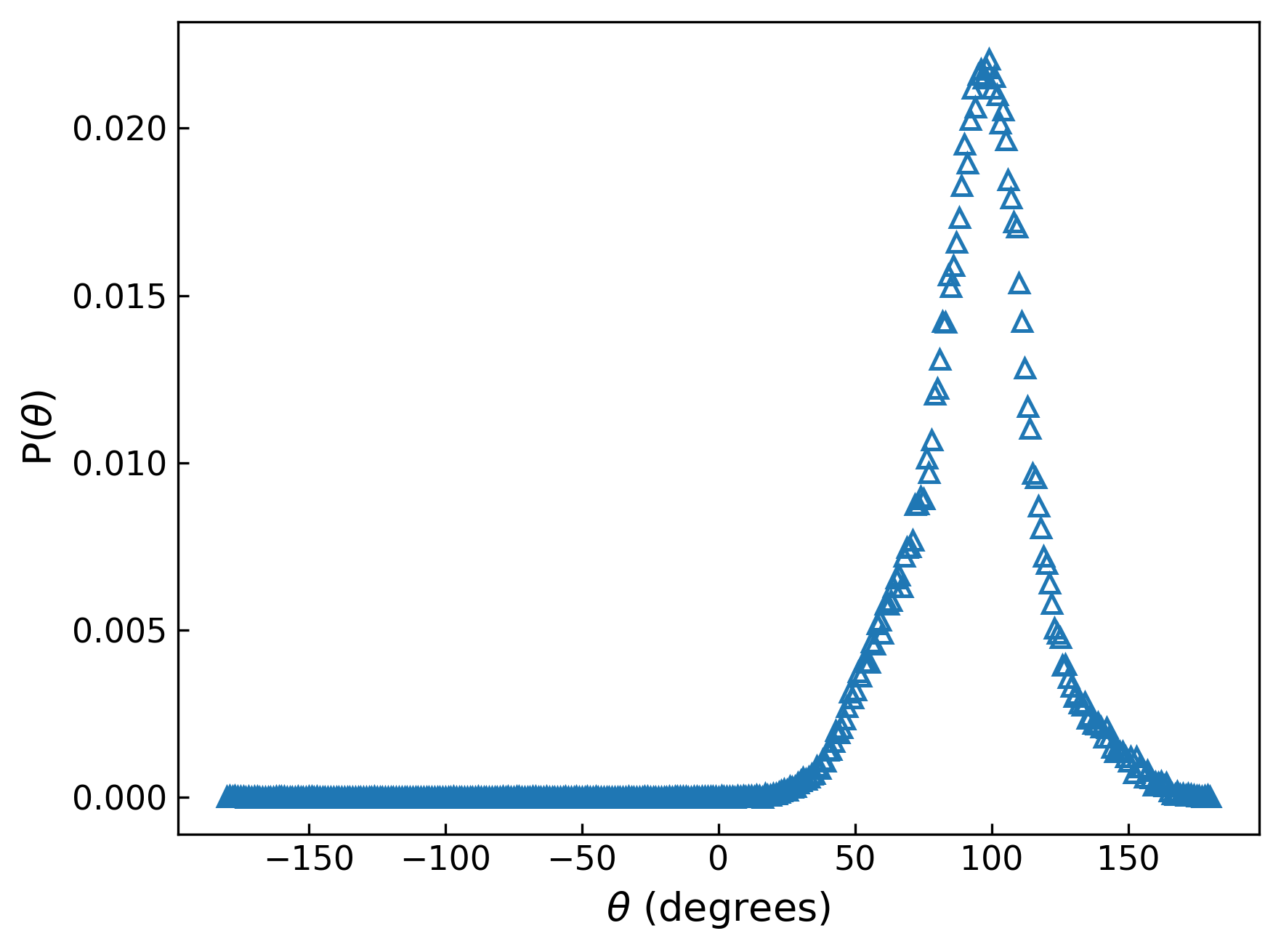}
     \caption{\textbf{AA dihedral-angle distribution at the amylose 1--4 glycosidic linkage.} Probability distribution (P ($\theta$)) of the torsion angle ($\theta$) at the $\alpha(1\rightarrow4)$ glycosidic linkage of amylose in the TPS melt at \SI{613}{\kelvin} and \SI{1}{\bar}, computed from the atomistic (AA) trajectory.} \label{fig:AAdiah}
 \end{figure*}

\begin{figure*}[!htb]
     \centering
     \includegraphics[width=15cm,height=13 cm]{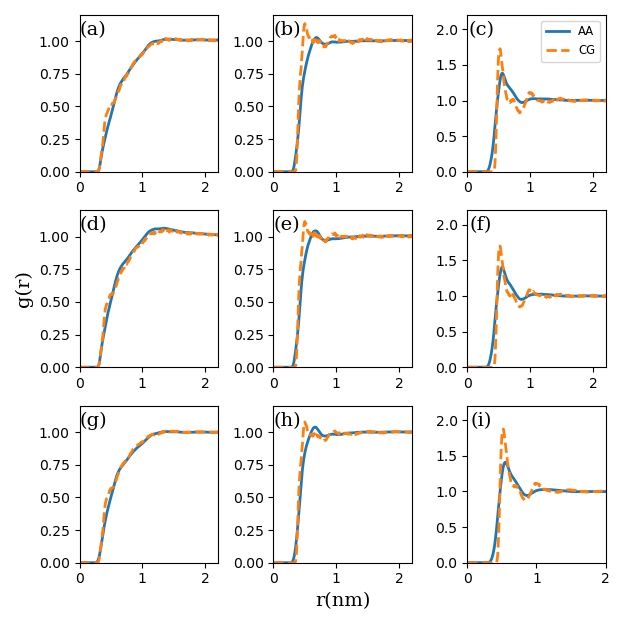}
     \caption{\textbf{Radial distribution function for polymer-sorbitol pairs in TPS-PY composite}. \textbf{(a)}, \textbf{(d)}, and \textbf{(g)} show RDF for polymer-polymer pairs; \textbf{(b)}, \textbf{(e)}, and \textbf{(h)} show RDF for polymer-sorbitol pairs; and \textbf{(c)}, \textbf{(f)}, and \textbf{(i)} show RDF for sorbitol-sorbitol pairs across the near-sheet, far-sheet, and melt. Data obtained from AA and CG of TPS melt and TPS-PY composite at \SI{613}{\kelvin} and \SI{1}{\bar}.}
     \label{fig:nearfarRDFpyrophyllite}
 \end{figure*}

MARTINI-3 polysaccharide model was used for amylose and amylopectin, parameterized using AA simulations of mono- and disaccharides in polar and apolar solvents \cite{lutsyk2022extending}. The CG model is based only on isotropic non-bonded interactions and cannot reproduce hydrogen-bonding-mediated helical structures of amylose or double-helical order in amylopectin. Under the melt-state conditions simulated here (\SI{613}{\kelvin}), AA simulations show the dihedral distribution at the amylose 1–4 linkage is broad and unimodal (Supplementary Information Figure \ref{fig:AAdiah}), with no signature of the narrow, discrete torsion states associated with helical amylose; branching in amylopectin further restricts double-helix formation at the chain lengths simulated here. This is consistent with experimental evidence that thermal and mechanical processing substantially disrupts native starch order: crystallinity fell from \SI{17.5}{\percent} to \SI{10.3}{\percent} under extrusion \cite{chen2022effect} and gelatinization enthalpy dropped to a small fraction of native values \cite{pushpadass2009macromolecular}. Omission of directional interactions nonetheless remains a limitation of the CG treatment, contributing — together with other effects of isotropic non-bonded interactions — to the $\mathrm{R_g}$ and RDF deviations from AA reported in this study. The RDFs obtained using AA and CG simulations of TPS melt had good agreement for polymer-polymer pair (Figure \ref{fig:nearfarRDFpyrophyllite}), with less than \SI{10}{\percent} error in the two-body excess entropy for both polymer-polymer and sorbitol-sorbitol pairs (Table \ref{tab:tbemelt}). For sorbitol-sorbitol pairs, AA and CG profiles show good agreement in peak positions, though the CG model overestimates first-shell peak heights by $\sim$ \SI{30}{\percent}, attributable to the isotropic nature of MARTINI LJ interactions. Additionally, an outward shift in the sorbitol-polymer RDF led to \SI{22.22}{\percent} higher $S_{2,\alpha,\beta}$ in the CG simulation (see Figure \ref{fig:nearfarRDFpyrophyllite}), suggesting the need for further improvement in cross-interaction parameters. However, the sorbitol-polymer pair contributes less than 10-12\SI {}{\percent} to overall $S_2$, making the developed TPS model still suitable for subsequent calculations. The CG simulation overestimated (w.r.t. AA) $R_\mathrm{g}$ by \SI{17.8}{\percent} and \SI{16.7}{\percent} for amylose and amylopectin, respectively, leading to underestimation of $S_\mathrm{c}$ by \SI{8}{\percent} and \SI{6.2}{\percent} respectively (Table \ref{tab:pyrophylliteRgScDn}), and overestimation of TPS density by \SI{5.1}{\percent} (\SI{1280.7}{\kg\per\cubic\meter} in CG). 
Inspite of these shortcomings, the present MARTINI-3 TPS model provides a better geometric representation of starch monomers, a better balance between polymer-sorbitol interactions, comparable chain-level properties, and a better TPS density compared to available MARTINI-2 models (data provided in Table \ref{tab:tbemelt}) \cite{Lopez2009}. Importantly, there was no requirement for scaling of specific cross- or self-interactions, which is expected to improve transferability of the potential. Large-scale AA simulations of long-chain amylose and amylopectin can be used to further refine the MARTINI-3 CG model for TPS \cite{patidar2024martini}. 
\begin{table*}[!htb]
 \small
  \centering
  \captionof{table}{\textbf{Comparison of properties obtained using AA and CG simulations of TPS melt and TPS-MMT-TMA composite.} The composite system CG simulations were run with $Mg$-connected $\mathrm{SiO_2}$ sites assigned various bead types in range $\ce{SN_6}$-$\ce{SP_6}$ and other $\mathrm{SiO_2}$ sites fixed at $\ce{SC_6}$. The average values of $R_\mathrm{g}$, $S_\mathrm{c}$, and $D_\mathrm{n}$ were calculated using last \SI{900}{\nano\second}  trajectories at \SI{613}{\kelvin} and \SI{1}{\bar}.}
  \centering
  \label{tab:TMSMMTRgScDn}
\begin{threeparttable}
\resizebox{0.99\textwidth}{!}{
  \begin{tabular}{ c c c c c c c c c c } 
    \toprule
    
      \multicolumn{3}{c}{} & \multicolumn{4}{c} {\textbf{Amylose}} & \multicolumn{3}{c}{\textbf{Amylopectin}}   \\
    \cmidrule(lr){4-6}\cmidrule(lr){7-10}
      \multicolumn{2}{c}{} &\multicolumn{1}{c}{\textbf{bead-type}} &\multicolumn{1}{c}{\textbf{near}}  & \multicolumn{1}{c}{\textbf{far}}  &  \multicolumn{1}{c}{\textbf{melt}}  &   \multicolumn{1}{c}{} & \multicolumn{1}{c}{\textbf{near}} & \multicolumn{1}{c}{\textbf{far}} & \multicolumn{1}{c}{\textbf{melt}} \\
    \midrule
    $R_\mathrm{g}$(\SI{}{\nano\meter}) &    AA & & 1.32$\pm$0.09    & 1.36$\pm$0.08   & 1.38$\pm$0.02  &  & 1.72$\pm$0.15 & 1.77$\pm$0.16 & 1.83$\pm$0.02\\
    & CG      &  &  &   & 1.65$\pm$0.04 &  &  &  & 2.16$\pm$0.08 \\
  
    &       & $\ce{SN_6}$ &1.59$\pm$0.11    & 1.64 $\pm$0.08  &  &  & 2.07$\pm$0.18 & 2.15 $\pm$0.12&  \\
    &       & $\ce{SP_1}$ & 1.61$\pm$0.09	&1.64 $\pm$0.07&	 && 2.13$\pm$0.16	& 2.12$\pm$0.13&	\\

    &      & $\ce{SP_2}$ & 1.55$\pm$0.12 & 1.64$\pm$0.10&  && 2.11$\pm$0.17 & 2.15$\pm$0.12	& \\
    &       & $\ce{SP_4}$ & 1.60$\pm$0.13 & 1.64$\pm$0.12& && 2.09$\pm$0.18&	2.14$\pm$0.14	& \\
    &       & $\ce{SP_6}$ & 1.59$\pm$0.11	& 1.61$\pm$0.09&	 && 2.08$\pm$0.17	&2.12$\pm$0.13	&\\

     \cmidrule(lr){3-10}
${\langle S_c \rangle}$  & AA  && 37.72$\pm$1.24  & 39.11$\pm$1.07 & 37.97$\pm$0.11  &  & 36.23$\pm$1.64 & 41.92$\pm$1.59 & 39.30$\pm$0.18 \\
$(\SI{}{\joule\per\mol\per\kelvin})$   & CG &     &   &  & 34.91 $\pm$0.14 &  &  &  & 36.87 $\pm$0.11\\
   &  & $\ce{SN_6}$     & 34.42$\pm$1.28    & 36.40 $\pm$1.19  &  &  & 36.01$\pm$1.75 & 40.01 $\pm$1.68&  \\
&&$\ce{SP_1}$&34.27$\pm$1.01&	36.20$\pm$1.22&	&& 36.75$\pm$1.80	&40.21$\pm$1.52	&\\
&&$\ce{SP_2}$&34.12$\pm$1.31&	36.29$\pm$1.14&	 && 36.02$\pm$1.62	& 39.35$\pm$1.70	& 
\\
&&$\ce{SP_4}$&34.14	$\pm$1.18 &36.53 $\pm$1.27&	&& 36.25$\pm$1.67&	39.54$\pm$1.64&
\\
&&$\ce{SP_6}$&34.83$\pm$1.30&	36.54$\pm$1.20&	 && 36.10$\pm$1.77	& 40.10$\pm$1.69	& 
\\
\midrule

$D_\mathrm{n}$  & AA  && 0.96$\pm$0.12  & - & -  &  & 1.16$\pm$0.14 & - & - \\ 
& CG & $\ce{SN_6}$     & 0.78$\pm$0.10    & -   & -  &  & 0.94$\pm$0.13 & - & -\\
&&$\ce{SP_1}$& 0.74$\pm$0.09 &	-&-&& 1.00$\pm$0.08	& -	& -\\
&&$\ce{SP_2}$& 0.74$\pm$0.11 &	-&-&& 0.79$\pm$0.10	& -	& -\\
&&$\ce{SP_4}$&0.67$\pm$0.14&	-&	-&& 0.79$\pm$0.15	& -	& -\\
&&$\ce{SP_6}$& 0.66	$\pm$0.17& -&-&& 0.84$\pm$0.18 &	-&	-\\
 \bottomrule
  \end{tabular}}
  \end{threeparttable}
\end{table*}

\subsubsection{Determination and Validation of CG Bead Types for Clays}\label{clay non-bonded parameters}
\begin{figure}[!htb]
     \centering
     \includegraphics[width=14 cm,height=8 cm]{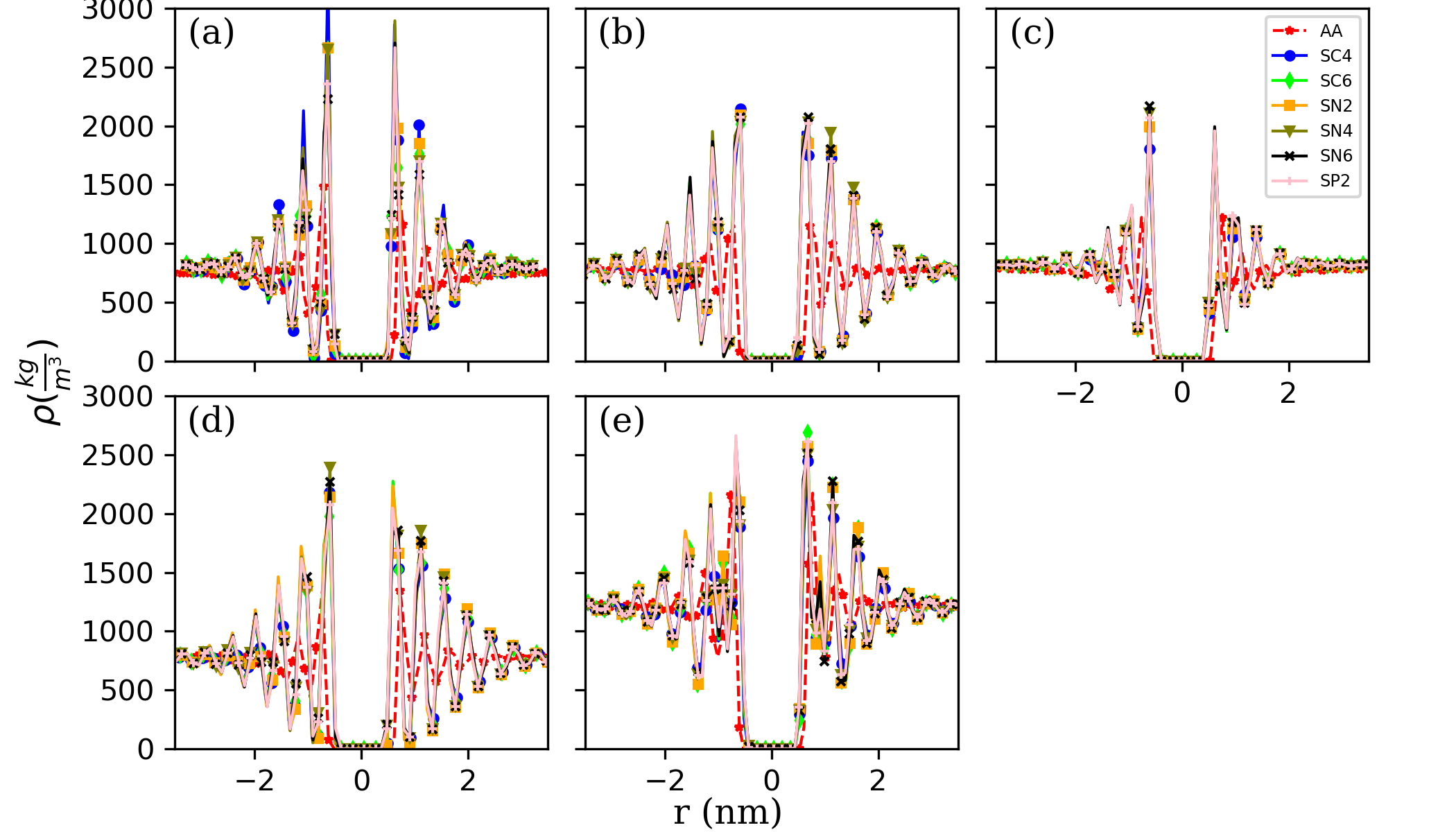}
     \caption{ \textbf{Density profile of small solvent molecules along the z-direction of pyrophyllite clay}. \textbf{(a)} Hexadecane, \textbf{(b)} propanol, \textbf{(c)} ethanol, and \textbf{(d)} iso-propanol density profile along the z-direction of pyrophyllite plane were estimated at \SI{300}{\kelvin} and \SI{1}{\bar}. \textbf{(d)} sorbitol density was estimated at \SI{450}{\kelvin} and \SI{1}{\bar}. The CG simulations were performed with different CG bead assignments for surface $\mathrm{SIO_2}$. The data is shown for SC4 (\bluewithcircle), SC6 (\greentriiangle), SN2 (\orangewithsquare), SN4 (\oliveinvertedtriangle), SN6 (\blackmultiline), and SP2 (\pinkplus) bead assignments, and were plotted along with AA (\redstar) results at the same operating conditions.}
     \label{fig:densitysolvent}
 \end{figure}

\begin{figure*}[!htb]
     \centering
     \includegraphics[width=12cm,height=8.5cm]{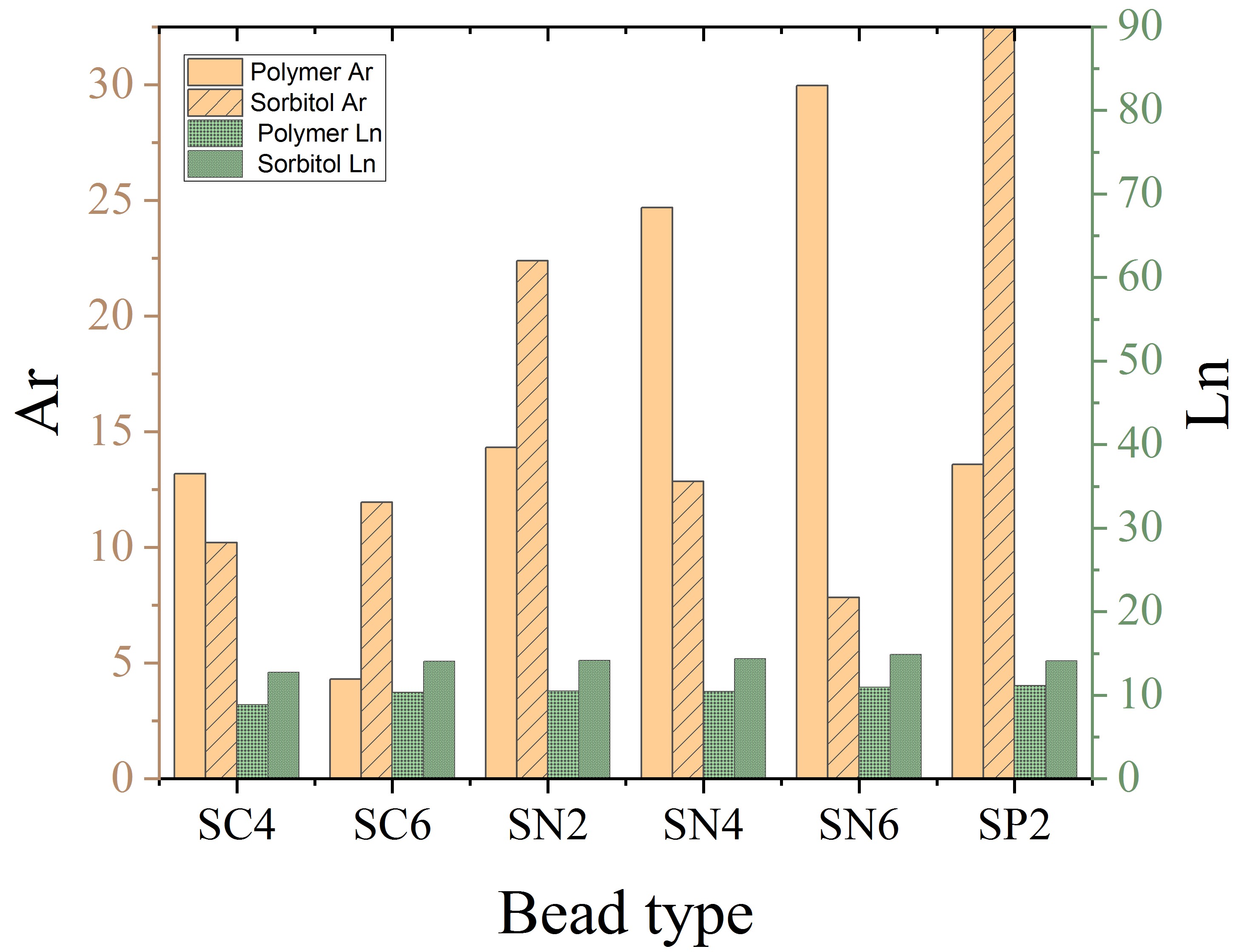}
     \caption{\textbf{Relative error in CG density profiles of TPS constituents for TPS-PY composite} The $\mathrm{Ar}$ and $\mathrm{Ln}$ were calculated using density profile integrals using Equations \ref{eq:Areq} and \ref{eq:Lneq}, respectively. The integration was performed till the end of the first solvation shell (\SI{1}{\nano\meter}). The left and right axes show relative percentage error in $\mathrm{Ar}$ and $\mathrm{Ln}$, respectively. The relative error in sorbitol $\mathrm{Ar}$ for $\ce{SP_2}$ bead was \SI{80.37}{\percent}.} \label{fig:pydensityerr}
 \end{figure*}

An initial estimate for clay bead types was obtained by comparing structural analogues among the functional groups with available MARTINI-3 assignments (Supplementary Tables 24 and 25 in Souza et al. \cite{souza2021martini}) and hydration energies (determined from experimental surface tension data) of clay silicates, which served as the starting point for iterative optimization using data from TPS-Clay simulations.\\ 
\paragraph{Pyrophyllite CG Parameters:}\label{pyrophyllite non-bonded parameters}
The octahedral $\ce{AlO_2H}$ groups were represented using $\ce{SN_{4a}}$ beads due to their similarity to acetal groups, while surface $\ce{SiO_2}$ groups were assigned to get the best possible match between the hydration energy of MARTINI-3 bead types (Supplementary Table 18 in Souza et al.\cite{souza2021martini}) and the calculated hydration energy of the pyrophyllite surface. The hydration energy was calculated by first determining the pyrophyllite surface tension ($\gamma_s$) components using the Lifshitz theory \cite{lifshitz1956theory} and the Young's Equation \ref{eq:surfacetension} \cite{vanoss1988additive}. The dispersive and polar component were calculated using contact angle and surface tension data of nonpolar ($\alpha$-bromonaphthalene and diiodomethane) and polar (water and ethylene glycol) liquids (data in Table \ref{tab:surface_energy} \cite{giese1991surface}), giving $\gamma_s^{LW}=\SI{34.4}{\milli\newton\per\meter}$, $\gamma_s^+ = \SI{3.2}{\milli\newton\per\meter}$ and $\gamma_s^- = \SI{1.7}{\milli\newton\per\meter}$.  
These surface tension components were used to calculate the hydration energy for the pyrophyllite surface as $\Delta F_\mathrm{{hyd, exp}} = \SI{4.62}{\kilo\joule\per\mol}$, using Equation \ref{eq:hydration} (values calculated per CG bead using its cross-sectional area, calculation details in Section \ref{surface hydration energies}). A previously reported linear relation between $\Delta F_\mathrm{{hyd, exp}}$ and hydration energies obtained using MARTINI-3 CG simulations ($\Delta F_\mathrm{{hyd, sim}}$) for 36 organic liquids (see Equation \ref{eq:extrapolationeq})\cite{souza2021martini} was used to get $\Delta F_\mathrm{{hyd, sim}}$ of \SI{-1.65}{\kilo\joule\per\mol} for pyrophyllite surface bead. The MARTINI-3 $\ce{SC_6}$ bead with $\Delta F_\mathrm{{hyd, sim}}=\SI{-2.00}{\kilo\joule\per\mol}$ represents the closest match for this pyrophyllite surface bead. The $\ce{SC_6}$ bead type is consistent with the MARTINI FF convention for representing non-polar, non-wetting surfaces, appropriately capturing the hydrophobic character \cite{giese1991surface} and low water-sorption energy (\SI{1}{\kilo\joule\per\mol}) of pyrophyllite surface \cite{churakov2006ab}.\\
Data from AA simulation of solvents with varying polarity, viz. hexadecane, propanol, ethanol, isopropanol, and sorbitol, on pyrophyllite basal (\{001\}) surface (sheet connected across periodic boundaries in the $x-y$ plane) was used to refine surface bead assignments. The density profiles along the sheet normal (Figure \ref{fig:densitysolvent}) and normalized diffusion coefficients ($D_\mathrm{n}$, see Table \ref{tab:solventDn}) obtained from CG simulations were nearly identical for different surface bead types for each solvent studied, indicating minimal sensitivity for properties of small molecule solvents. In contrast, polymer dynamics and density profiles showed a significant dependence on the surface bead type. Increasing the surface polarity beyond $\ce{SN_2}$ resulted in the freezing of polymer segments near the surface (Table \ref{tab:pyrophylliteRgScDn}) and high relative errors in density profile integrals ($Ar$) for both polymer and sorbitol (Figure \ref{fig:pydensityerr}). The relative errors for polymer and sorbitol $Ar$ were \SI{13.17}{\percent} and \SI{10.21}{\percent} for $\ce{SC_4}$, \SI{4.31}{\percent} and \SI{11.95}{\percent} for $\ce{SC_6}$, and \SI{14.65}{\percent} and \SI{22.51}{\percent} for $\ce{SN_2}$ (see Figure \ref{fig:pydensityerr}). The relative error in the average distance of a TPS bead from surface, $Ln$, was \SIrange{8}{12}{\percent} for different bead choices. The average (over amylose and amylopectin) relative percentage error in the near-to-far $R_\mathrm{g}$ and $S_\mathrm{c}$ ratios was \SI{4.8}{\percent} and \SI{1.75}{\percent} for $\ce{SC_4}$, \SI{5.02}{\percent} and \SI{1.06}{\percent} for $\ce{SC_6}$, and \SI{6.21}{\percent} and \SI{3.26}{\percent} for $\ce{SN_2}$, respectively. The overall results indicated $\ce{SC_4}$ and $\ce{SC_6}$ as two suitable bead types for surface $\ce{SIO2}$. The latter matches the initial bead assignment done using the hydration free energy estimates, and was thus used for all further calculations. Wide availability of experimental surface tension data will allow application of this approach for the parameterization of other inorganic functional groups. The SASA for pyrophyllite clay (TPS-PY system) was obtained to be 223.29 $\pm$ 0.05 \SI{}{\nano\meter\squared} and 225.98 $\pm$ 0.09 \SI{}{\nano\meter\squared} in AA and CG simulations, respectively, showing an excellent agreement. This result confirms that the CG mapping preserves the correct lateral dimensions and exposed surface area of the pyrophyllite platelet relative to the underlying atomistic representation. The normalized density distributions ($\rho_\mathrm{n} = \rho_\mathrm{composite}(z)/\rho_\mathrm{melt}$) for the polymer and sorbitol obtained using $\ce{SC_6}$ surface bead showed good agreement between AA and CG (Figure \ref{fig:de}). However, a lower water density near the pyrophyllite surface for CG systems shows potential incompatibility with the MARTINI-3 water model. Furthermore, $D_\mathrm{n}$ for small molecule organic solvents and TPS components was underestimated in CG simulations (w.r.t. AA) by $\sim35$-50\SI{}{\percent}. This indicates a limitation of the MARTINI-3 FF in accurately capturing dynamics at the solid-fluid interface, but a high accuracy for several structural and thermodynamic properties is expected to allow wider utility. Further, the use of MARTINI-3 FF allowed overcoming one major limitation of the unphysical freezing of solvents near solid interfaces in previous MARTINI FF models. Although, this was addressed by downscaling of surface-solvent interactions in older MARTINI FF models \cite{gobbo2013martini, patidar2024martini}, it limited parameter transferability to diverse chemical environments. 

\begin{figure*}[!htb]
     \centering
     \includegraphics[width=12cm,height=8.5cm]{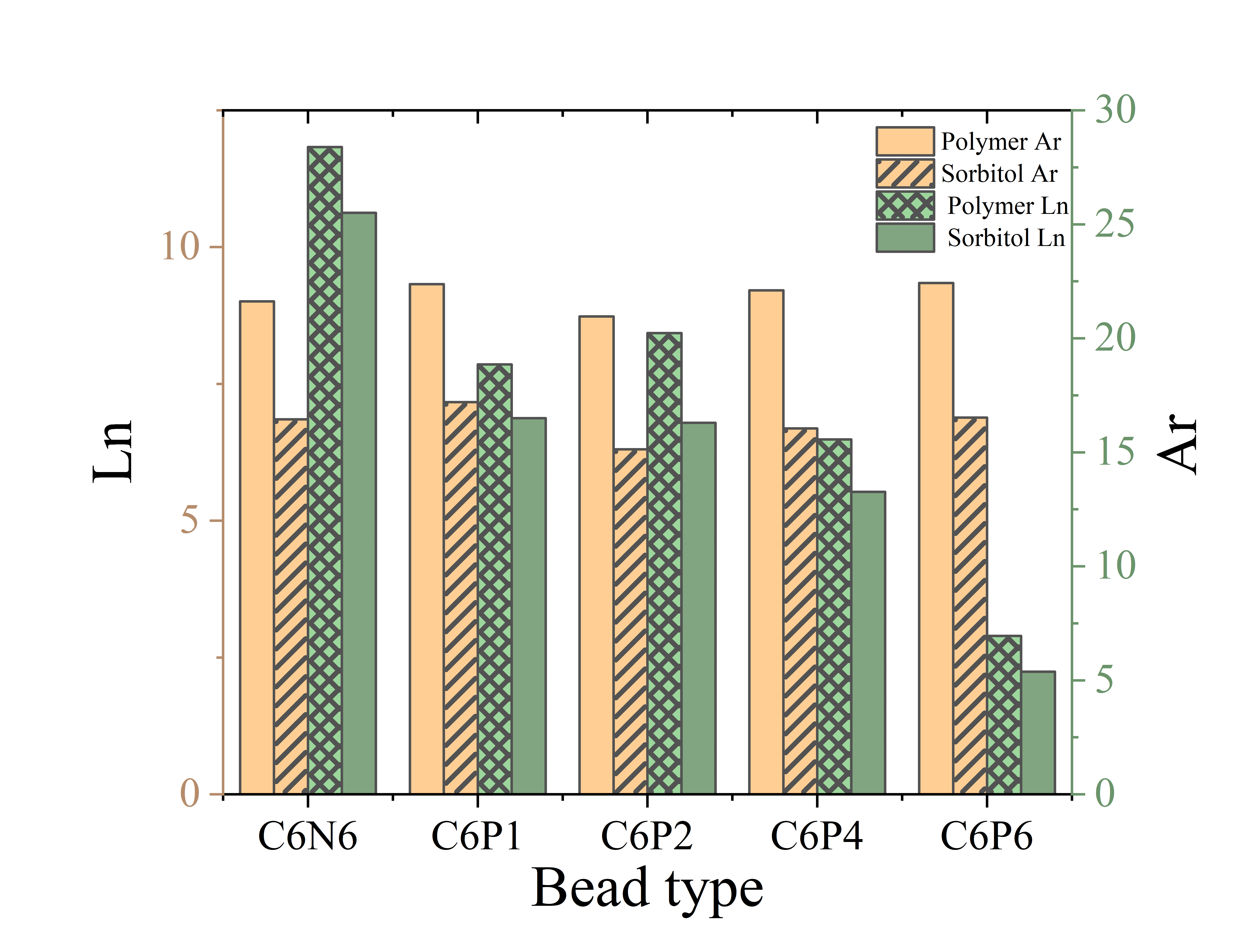}
     \caption{\textbf{Relative error in CG density profiles of TPS constituents for TPS--TMA-MMT composite} The $\mathrm{Ar}$ and $\mathrm{Ln}$ were calculated using Equations \ref{eq:Areq} and \ref{eq:Lneq}, respectively. The integration was performed till the end of the first solvation shell (\SI{1}{\nano\meter}). left and right axes show relative percentage error in $\mathrm{Ln}$ and $\mathrm{Ar}$, respectively.} \label{fig:mmtdensityerr}
 \end{figure*}

\paragraph{TMA-MMT CG Parameters:}\label{TMA-MMT non-bonded parameters}
The $\ce{MgO_2H}$ unit ($-1 \,e$ charge) obtained by isomorphic substitution of $\ce{Al^{3+}}$ with $\ce{Mg^{2+}}$ in TMA-MMT octahedral layer was modeled by a $\ce{SQ_5}$ bead due to its similarity with the phosphate ion (lipid head). The intergallery TMA counterions ($+1 \,e$ charge) were represented by a $\ce{Q_2}$ bead, as per available MARTINI-3 assignment \cite{souza2021martini}. Modeling all surface $\ce{SiO_2}$ beads by a $\ce{SC_6}$ bead (optimal assignment for pyrophyllite) will not account for high local polarity induced by highly electronegative oxygen atoms bonded to octahedral $\ce{Mg}$, as inferred by a lower sorbitol density near the surface. This finding aligns with iso-surface electrostatic potential calculations showing negative surface charges are concentrated on surface oxygen atoms in the central unit cell region of $\ce{Mg}$ substitution sites \cite{lavikainen2015montmorillonite}. Therefore, we modeled MMT surface $\mathrm{SiO_2}$ with $\ce{SC_6}$ beads for $\ce{Al}$-connected sites and tested several higher-polarity bead types, viz. $\ce{SN_6}$, $\ce{SP_1}$, $\ce{SP_2}$, $\ce{SP_4}$, and $\ce{SP_6}$, for $\ce{Mg}$-connected $\mathrm{SiO_2}$ sites (one-third of total surface beads). Accuracy in dynamics and density distributions showed a tradeoff: relative error in polymer $D_\mathrm{n}$ (averaged over amylose and amylopectin trimeric segments) increased from \SI{11.29}{\percent} to \SI{29.20}{\percent} (Table \ref{tab:TMSMMTRgScDn}), while polymer and sorbitol $\mathrm{Ar}$ errors decreased from \SI{28.39}{\percent} to \SI{6.94}{\percent} and from \SI{25.50}{\percent} to \SI{5.37}{\percent}, respectively (Figure \ref{fig:mmtdensityerr}). The relative error in $\mathrm{Ln}$ for both pairs was less than $\sim$\SI{10}{\percent} across all assignments. The $\ce{SP_1}$ bead, which gave intermediate deviations of \SI{16.49}{\percent} for sorbitol $Ar$ and \SI{18.61}{\percent} for polymer $\ce{D_n}$, was chosen for these higher polarity surface beads. The CG predictions for normalized $R_\mathrm{g}$ and $S_\mathrm{c}$ (near-surface to melt ratio) were within \SI{7}{\percent} of AA for all CG assignments (see Table \ref{tab:TMSMMTRgScDn}). Therefore, the simple two-bead type model for surface $\mathrm{SiO_2}$ represents an acceptable CG model for TMA-MMT for use with both small organic solvents as well as polymer matrices.

\paragraph{Edge bead selection for TMA-MMT:} Phyllosilicates exhibit distinct cleavage patterns leading to edge surfaces such as \{110\}, \{010\}, \{101\}, and \{130\}. Here, we have modeled the most stable \{010\} and \{100\} surfaces, which comprise of tetrahedral layer $\mathrm{SiOH}$ and octahedral layer $\ce{Al}$ with one (at \{010\} surface) or two (at \{100\} surface) hydroxyl groups, respectively. The \{010\} $\ce{Al}$ sites were assigned the $\ce{P_2}$ bead type due to their structural similarity to the hemiacetal groups (single $\ce{Al-OH}$ bond), while \{100\} $\ce{Al}$ sites were assigned $\ce{P_4}$ as they resembled the diol groups (two $\ce{Al-OH}$ bonds). The $\ce{SiOH}$ groups were represented using $\ce{P2}$ beads, as the resulting silanol functional groups also closely mimic a hemiacetal group. The $\mathrm{MgOH_2}$ groups were mapped to $\ce{SQ_5}$, regardless of their position (edge or interior). These assignments accurately represented functional groups according to the MARTINI-3 framework, while preserving relative polarity differences between edge types \cite{souza2021martini}. For instance, the \{010\} edge has a lower sorption energy of -\SI{38.9}{\kilo\joule\per\mol} and was represented by $\ce{P_2}$ bead, whereas the \{100\} edge has a higher sorption energy of -\SI{57.7}{\kilo\joule\per\mol} and was represented by $\ce{P_4}$ beads \cite{churakov2006ab}. The $\ce{P2}$ bead type for the \{010\} edge was previously shown to accurately reproduce the cleavage free energy for a clay dimer in contact at this surface. \cite{khan2025characterization}. 

\paragraph{Additional Validation of Developed CG Parameters:}

  \begin{table}[!htb]
 \small
 \centering
  \caption{ \textbf{Normalized two-body excess entropy for TPS-clay composite systems.} Two-body excess entropies (${S_{2,\alpha,\beta}}$) of all polymer-sorbitol pairs within TPS-PY and TPS--TMA-MMT systems were calculated using their respective radial distribution functions (RDFs), as shown in Figures \ref{fig:nearfarRDFpyrophyllite} and \ref{fig:nearfarRDFTMAMMT}. Values are reported here after dividing by the corresponding ${S_{2,\alpha,\beta}}$ in TPS melt (Table \ref{tab:tbemelt}). Five bonded atoms were excluded from all RDF calculations. P and S refer to polymer (amylose and amylopectin) and sorbitol beads, respectively.}
  \label{tab:tbe}
      \begin{threeparttable}
\resizebox{0.5\textwidth}{!}{
  \begin{tabular}{c c c c c c c c}
    \toprule
      \multicolumn{2}{c}{System} & \multicolumn{3}{c}{near} & \multicolumn{3}{c}{far}   \\
    \cmidrule(lr){3-5}\cmidrule(lr){6-8}
      \multicolumn{2}{c}{} & \multicolumn{1}{c}{P-P}  &  \multicolumn{1}{c}{P-S}  &  \multicolumn{1}{c}{S-S}  & \multicolumn{1}{c}{P-P}  &  \multicolumn{1}{c}{P-S} & \multicolumn{1}{c}{S-S}\\
    \midrule
         \multicolumn{3}{l}{TPS-PY} & \multicolumn{2}{c}{} & \multicolumn{3}{c}{}   \\
    \midrule
    AA &           & 1.21 $\pm$ 0.04  & 1.44 $\pm$ 0.06 & 0.98 $\pm$ 0.03 & 0.95 $\pm$ 0.03 & 1.30 $\pm$ 0.05 & 1.14 $\pm$ 0.04  \\
    CG &      & 1.17 $\pm$ 0.06   & 1.33 $\pm$ 0.07 & 0.92 $\pm$ 0.04 & 0.94 $\pm$ 0.03 & 1.23 $\pm$ 0.04& 0.88 $\pm$ 0.02  \\
   \midrule
         \multicolumn{4}{l}{TPS--TMA-MMT} & \multicolumn{2}{c}{} & \multicolumn{2}{c}{}   \\
    \midrule
     AA  &   & 1.29 $\pm$ 0.05   & 1.40 $\pm$ 0.05 & 1.02 $\pm$ 0.03 & 0.91 $\pm$ 0.02& 1.24 $\pm$ 0.04 & 1.05 $\pm$ 0.03  \\
    CG &       & 1.24 $\pm$ 0.06   & 1.31 $\pm$ 0.04 & 0.97 $\pm$ 0.05 & 0.98 $\pm$ 0.02 & 1.25 $\pm$ 0.03 & 0.88 $\pm$ 0.02  \\
    \bottomrule
  \end{tabular}}
\end{threeparttable}
\end{table}

\begin{figure*}[!htb]
     \centering
     \includegraphics[width=15cm,height=13 cm]{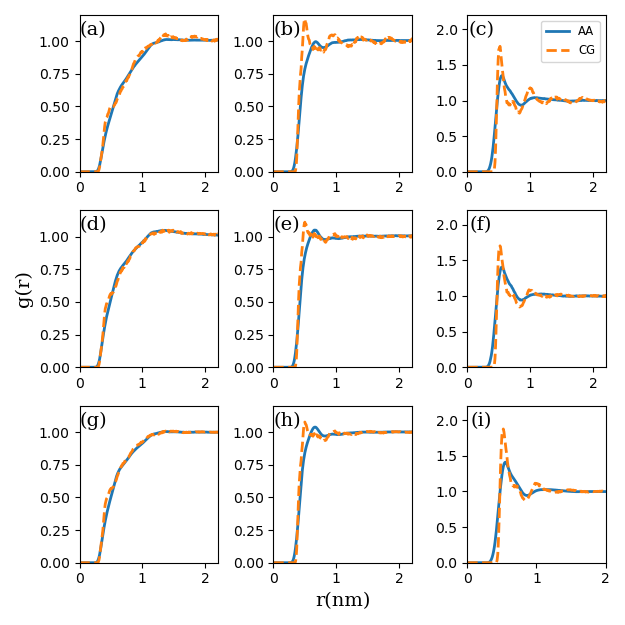}
     \caption{\textbf{Radial distribution function for polymer-sorbitol pairs in TPS-TMA--MMT composite}. \textbf{(a)}, \textbf{(d)}, and \textbf{(g)} show RDF for polymer-polymer pairs; \textbf{(b)}, \textbf{(e)}, and \textbf{(h)} show RDF for polymer-sorbitol pairs; and \textbf{(c)}, \textbf{(f)}, and \textbf{(i)} show RDF for sorbitol-sorbitol pairs across the near-sheet, far-sheet, and melt. Data obtained from AA and CG of TPS melt and TPS-TMA--MMT composite at \SI{613}{\kelvin} and \SI{1}{\bar}.}
     \label{fig:nearfarRDFTMAMMT}
 \end{figure*}

There was an excellent agreement between AA and CG RDFs (Figures \ref{fig:nearfarRDFpyrophyllite} and \ref{fig:nearfarRDFTMAMMT}) and pairwise ${S_{2,\alpha,\beta}}$ values for TPS-PY and TPS--TMA-MMT systems (Table \ref{tab:tbe}). The $S_{2,\alpha,\beta}$ (calculated from RDFs) for polymer-polymer (P-P), polymer-sorbitol (P-S), and sorbitol-sorbitol (S-S) pairs in the near region had errors of \SI{3.25}{\percent}, \SI{7.50}{\percent}, and \SI{5.50}{\percent}, respectively, in the TPS-PY system and \SI{4.05}{\percent}, \SI{6.52}{\percent}, and \SI{5.24}{\percent}, respectively, in the TMA-MMT system. Since RDFs were not used in CG parametrization, the high accuracy in these predictions establishes the robustness of the CG parameters developed here for high-resolution representation of investigated systems. 
\subsection{Structural and Thermodynamic Properties of TPS-PE and TPS-PE-CL Composite}

     \begin{figure*}[!htb]
     \centering
     \includegraphics[width=17cm,height=7cm]{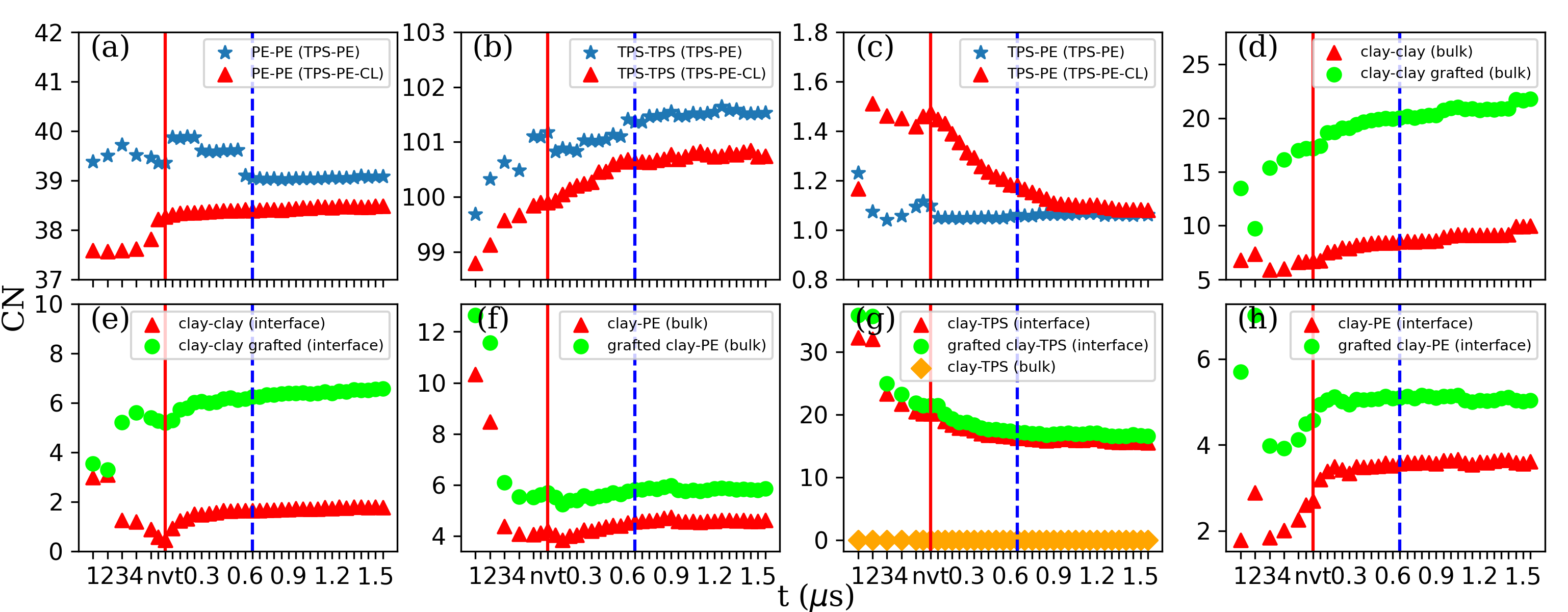}
     \caption{ \textbf{Morphology Evolution in TPS-PE melt and TPS-PE-CL composite.} The time evolution of second shell coordination numbers (CNs) in TPS-PE ($\bluestar$) and TPS-PE-CL ($\redtriiangle$, $\greencircle$, $\orangediamond$) systems was calculated from RDFs obtained from CG simulations at \SI{613}{\kelvin} and \SI{1}{\bar}  (Figure \ref{fig:RDFTPSPECL}). CN for: \textbf{(a)} PE-PE, \textbf{(b)} TPS-TPS, \textbf{(c)} TPS-PE, \textbf{(d)} clay-clay (in PE "bulk" phase), \textbf{(e)} clay-clay (at the TPS-PE "interface"), \textbf{(f)} clay-PE (bulk), \textbf{(g)} clay-TPS (interface), and \textbf{(h)} clay-PE (interface). The $x$-axis is partitioned to mark the following: T-P annealing cycles 1-4, end of \SI{25}{\nano\second} NVT simulation at \SI{613}{\kelvin} ($\redline$), NPT trajectory at \SI{613}{\kelvin} and \SI{1}{\bar} used for average property calculation ($\bluedotline$).}
     \label{fig:CN}
 \end{figure*}
 
\begin{figure}[!htb]
     \centering
     \includegraphics[width=17.65cm,height=6cm]{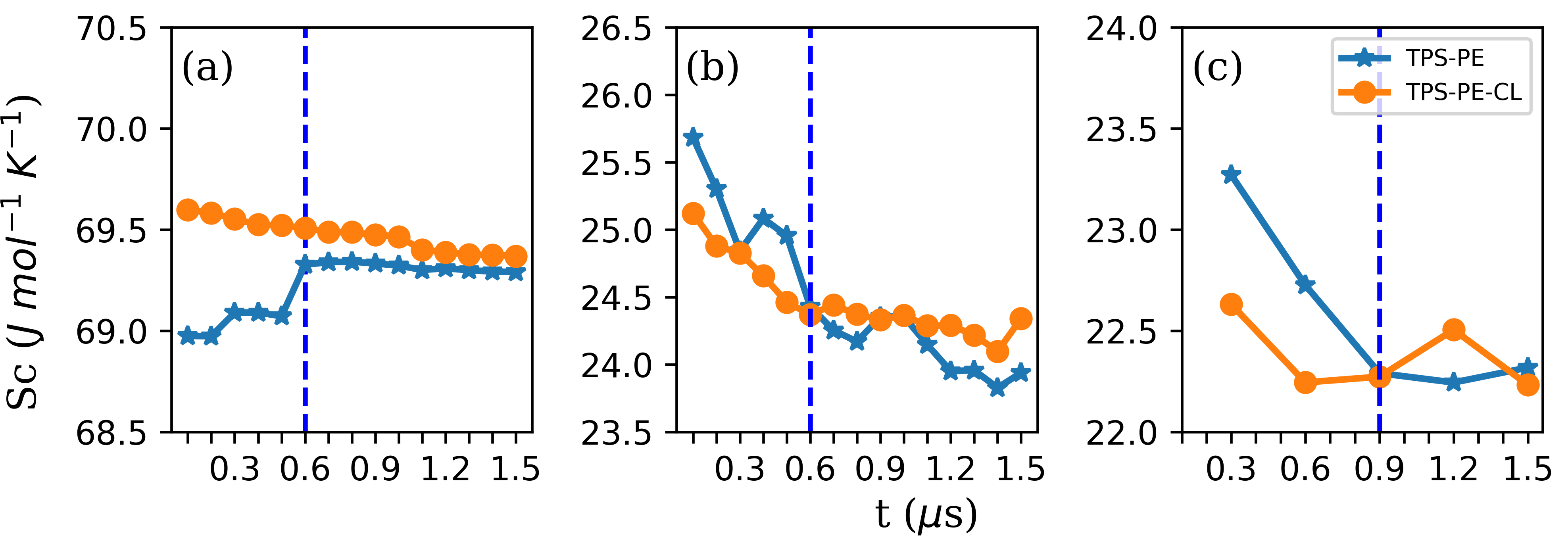}
     \caption{ \textbf{Temporal Profile polymer conformational entropy per CG bead in TPS-PE melt and TPS-PE-CL composite system}. \textbf{(a)} The $S_\mathrm{c}$ of polyethylene, and \textbf{(b)} amylose was calculated in blocks of \SI{100}{\nano\second}. \textbf{(c)} A larger block size of \SI{300}{\nano\second} was used for amylopectin. All data are calculated using CG MARTINI simulation of TPS-PE (\bluestar) and TPS-PE-CL (\orangewithcircle) systems at \SI{613}{\kelvin} and \SI{1}{\bar}. The $\bluedotline$ represents the end of the equilibration period, and data points following this line were used to calculate the average and error values reported in Table \ref{tab:conformation}.}
     \label{fig:Sc}
 \end{figure}

Temporal evolution of CN, $R_\mathrm{g}$, and $S_\mathrm{c}$ was used to determine convergence of structural and thermodynamic properties in CG simulations of TPS-PE and TPS-PE-CL systems. We observed significant changes in all properties during $T$-$P$ annealing cycles indicated extensive configurational sampling, followed by a gradual change and emergence of a plateau near $\sim\SI{600}{\nano\second}$ (averages calculated from the final \SI{900}{\nano\second} trajectory) in a long NPT simulation at \SI{613}{\kelvin} and \SI{1}{\bar} (see Figures \ref{fig:CN}, \ref{fig:Sc}, and \ref{fig:Rg}). For a few properties, the temporal profiles continued to evolve, albeit very slowly, till the end of the simulation, indicative of slow dynamics in these polymer composite systems. For these, we used the final \SI{600}{\nano\second} trajectory for calculation of averages, wherein the largest variation (for $R_\mathrm{g}$ of PE chains) over this period  was less than \SI{3}{\percent}.
 
\subsubsection{Morphology Evolution}
\begin{figure*}[!htb]
     \centering
     \includegraphics[width=15cm,height=10cm]{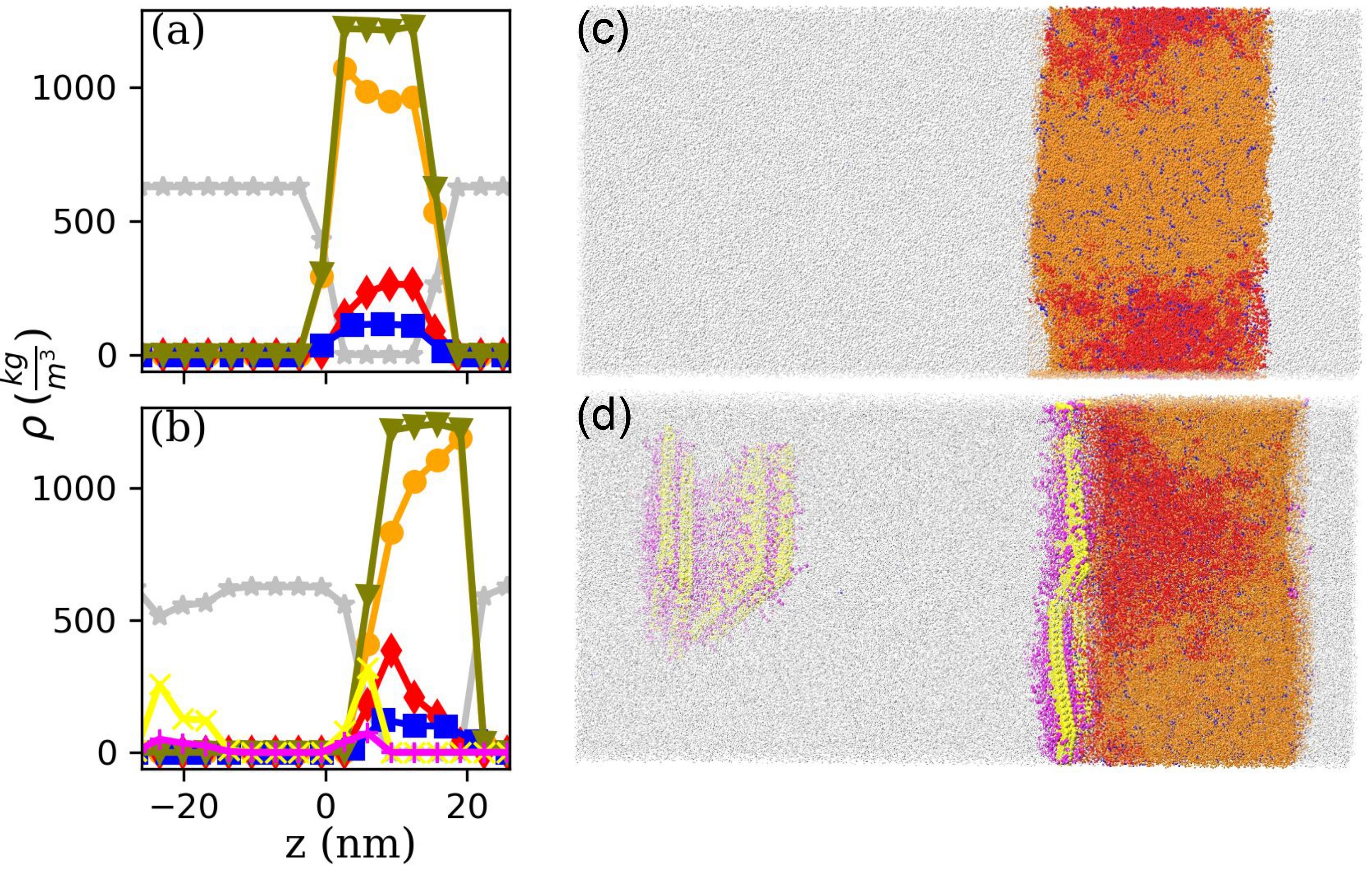}
     \caption{ \textbf{The density profile of TPS-PE and TPS-PE-CL constituents along the $z$-axis}. \textbf{(a)} Density profile of amylose (\reddiamond), amylopectin (\orangewithcircle), polysaccharide (\oliveinvertedtriangle), sorbitol (\bluewithsquare), PE (\graystar), sheet (\yellowmultiline), and grafts (\pinkplus) in TPS-PE system. \textbf{(b)} Density profile of the same components with additional graft and clay densities in the TPS-PE system. \textbf{(c)} TPS-PE melt and \textbf{(d)} TPS-PE-CL composite configurations at the end of the \SI{1500}{\nano\second} simulation. The color scheme in the snapshot matched their density profiles. All $\rho$ were calculated from the CG MARTINI simulation performed at \SI{613}{\kelvin} and \SI{1}{\bar}.}
     \label{fig:Densitylargesystem}
 \end{figure*}

\begin{figure}[!htb]
     \centering
     \includegraphics[width=7 cm,height=8cm]{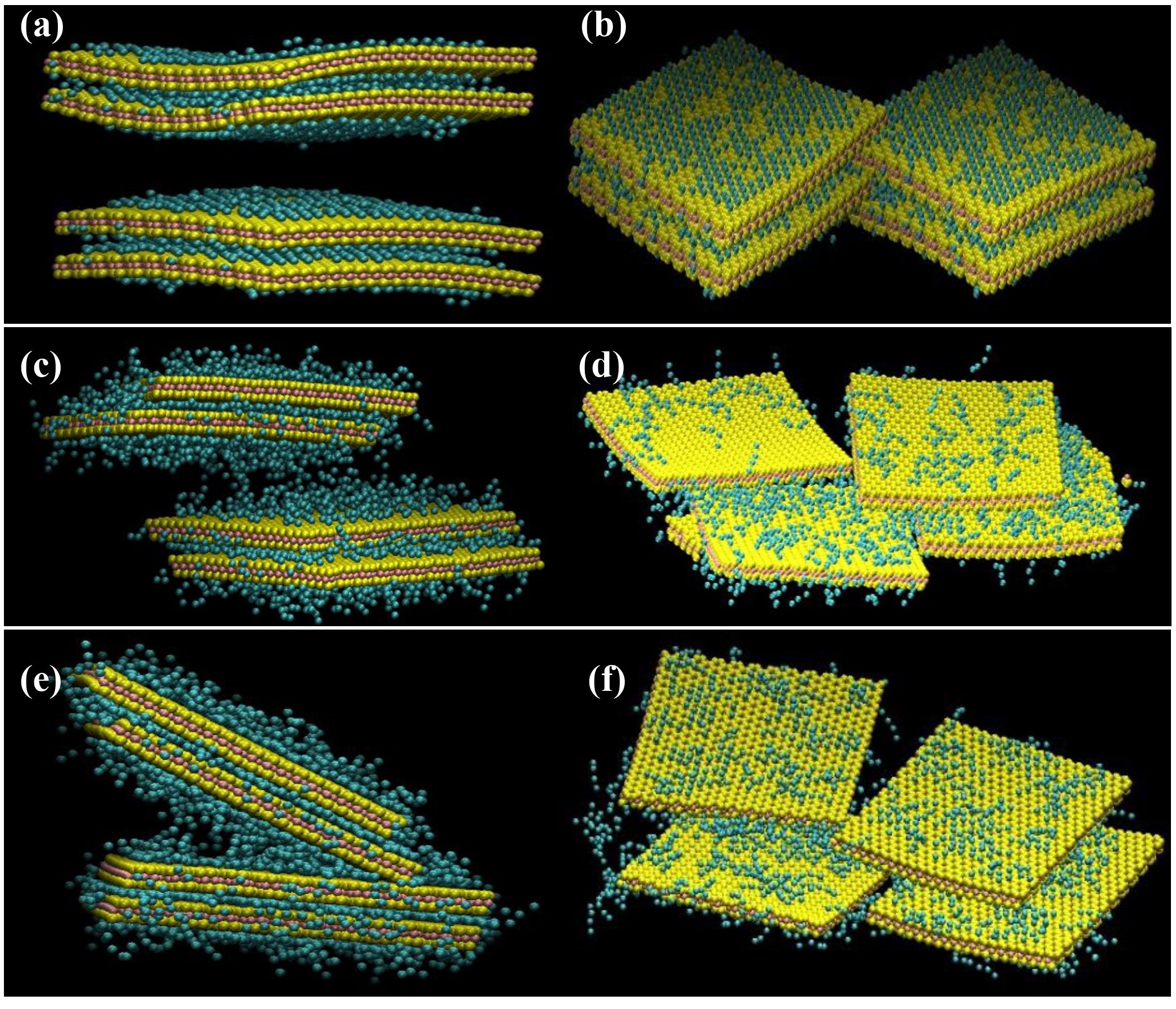}
     \caption{ \textbf{Time evolution of C12TAB-MMT clay tactoids in TPS-PE-CL composite system.} \textbf{(a)} Packmol-generated structure in PE bulk, \textbf{(b)} Packmol-generated structure at TPS-PE interface, \textbf{(c)} equilibrated structure in PE bulk after local equilibration, \textbf{(d)} equilibrated structure at TPS-PE interface after local equilibration, \textbf{(e)} final structures in PE bulk after \SI{1500}{\nano\second}, and \textbf{(f)} final structures at TPS-PE interface after \SI{1500}{\nano\second}. The alkyl grafts, surface ($\ce{SC_6}$ and $\ce{SP_2}$) beads, and the internal ($\ce{SQ_5}$ and $\ce{SN_{4a}}$) beads are represented by cyan, yellow, and pink colors, respectively.}
     \label{fig:sheetsnapshots}
 \end{figure}

\begin{figure*}[!htb]
     \centering
     \includegraphics[width=12cm,height=8cm]{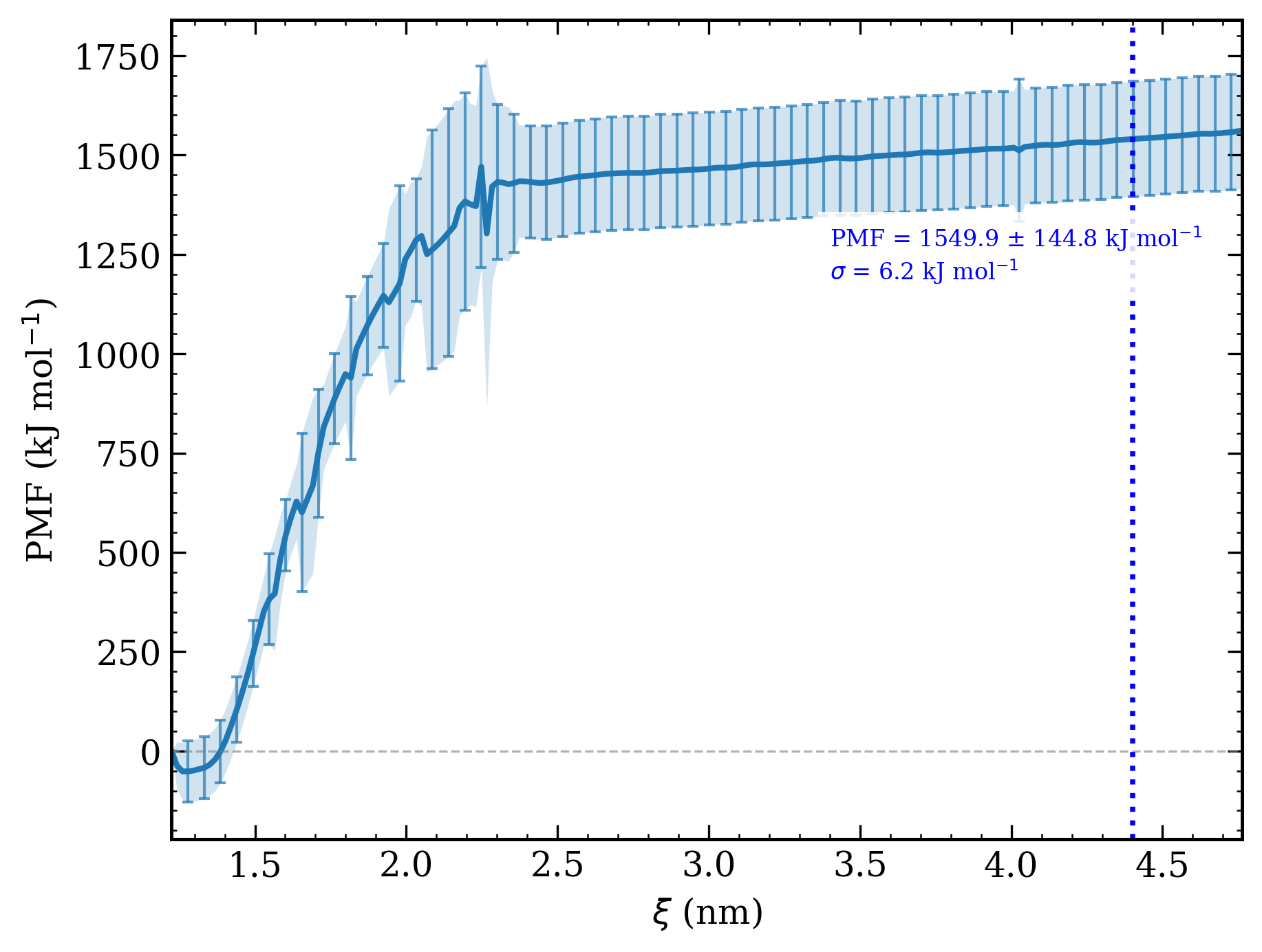}
\caption{\textbf{Potential of Mean Force (PMF) for the separation of a C12TAB-MMT clay sheet pair in a PE matrix.} The PMF was calculated using umbrella sampling with 38 windows spanning a reaction coordinate range of $\xi = 1.3$--$5.0$ \SI{}{\nano\meter} at intervals of $\Delta\xi = 0.1$ \SI{}{\nano\meter} and a temperature of \SI{613}{\kelvin}. Distributed four-point harmonic restraints were applied to maintain the target sheet separation in each window, and the PMF profile was reconstructed using the weighted histogram analysis method (WHAM). Initial configurations for the umbrella sampling windows were obtained from a constrained steered molecular dynamics (SMD) pulling trajectory. Vertical (\bluedotline) mark the point using for calculation of error and standard deviation ($\sigma$) in PMF.}
     \label{fig:pmf}
 \end{figure*}

Figures \ref{fig:CN} (a),(b) show that TPS-TPS CN converges to a value larger than PE-PE CN, as expected from the higher density for the TPS phase (Figure \ref{fig:Densitylargesystem} and refs. \cite{Ozeren2020a, panizon2015martini}). Both show a small decrease (1-3\SI{}{\percent}) in the composite system (w.r.t. melt), resulting from the formation of clay-polymer contacts. Further, their chemical incompatibility leads to a sharp transition in TPS and PE phase densities in the TPS-PE melt system (Figure \ref{fig:Densitylargesystem}(a)). The placement of C12TAB-MMT clay at the TPS-PE interface did not cause a noticeable change in TPS-PE CN (Figure \ref{fig:CN}(c)), but led to changes in the density profile of individual phases (Figure \ref{fig:Densitylargesystem}(b), discussed later) driven by interfacial localization of amphiphillic C12TAB-MMT. Figure \ref{fig:CN}(g) shows clay-TPS CN is essentially unchanged on inclusion of C12TAB grafts, implying TPS components bind predominantly to clay's polar surface. In contrast, a distinct increase in clay-PE CN on inclusion of C12TAB grafts can be attributed to an increased role of favorable hydrophobic interactions with the PE phase (Figures \ref{fig:CN}(f),(h)). 
The clay-TPS CN remains substantially higher than clay-clay and clay-PE CNs throughout the simulation. This interfacial localization of clay with preferential solvation in the TPS phase agrees well with experimental TEM micrographs of similar clay-polymer composites \cite{Sabetzadeh2016,sharif2011study}. Alkyl grafts contribute $\approx $\SI{70}{\percent} to clay-clay CN (Figures \ref{fig:CN} (d), (e)), and this high surface coverage, combined with their chemical similarity to PE, is expected to lead to high clay-PE CNs. For the interfacially localized clay platelets, the clay-clay (Figures \ref{fig:CN}(e)) and clay-PE (Figures \ref{fig:CN} (h)) CNs indeed converge to similar values of 6.44 and 5.07, respectively. However, for the clay platelets in the bulk PE phase, the clay-clay CN (Figures \ref{fig:CN}(d)) of 20.90 is almost 3x larger than the corresponding clay-PE CN. This is primarily driven by the interaction and efficient packing of short alkyl grafts (as evidenced by their significant contribution to clay-clay CN) combined with a weak quadropolar interactions of MMT clay platelets. These tactoid assemblies, inferred from bimodal clay density distribution in PE bulk (Figure \ref{fig:Densitylargesystem} (b) and representative snapshots (Figure \ref{fig:sheetsnapshots}) are also consistent with experimental observations in similar clay-polymer composites \cite{Sabetzadeh2016,sharif2011study}. To directly assess the thermodynamic stability of individual clay sheet pairs within these tactoids, we computed the PMF for separation of two C12TAB-MMT sheets in PE at \SI{613}{\kelvin} (Supplementary Information Figure \ref{fig:pmf}). The profile reveals a deep free energy minimum at $\sim$\SI{1.3}{\nano\meter}, rising steeply by $\sim$ \SI{1500}{\kilo\joule\per\mol} over 1.4--2.1\SI{}{\nano\meter} with no secondary minimum, confirming the complete absence of any partially-exfoliated metastable state. The large work of adhesion ($W_{\mathrm{adh}}$) of  \SI{5.68}{\milli\joule\per\meter\squared} at the clay–clay interface and no metastable intermediates establishes that tactoid formation is both kinetically and thermodynamically stable at \SI{613}{\kelvin}. This value is substantially lower than the interfacial surface tension of pristine pyrophyllite clay in air at \SI{298}{\kelvin} (\SI{34}{\milli\joule\per\meter\squared}), which can be attributed to the grafting of long alkyl chains onto the clay surface and the presence of the polyethylene (PE) matrix.

Further, density profiles reveal that each TPS component (amylose, amylopectin, and sorbitol) exhibits distinct interfacial behavior in melt and composite systems (Figure \ref{fig:Densitylargesystem}). In the TPS-PE melt, linear amylose chains experience a stronger PE repulsion than branched amylopectin, resulting in preferential amylopectin accumulation at the interface (Figures \ref{fig:Densitylargesystem} (a), (c)). In the composite system, however, amylose preferentially adsorbs to the clay surface, essentially displacing amylopectin at the TPS-PE interface compatibilized by clay platelets (Figures \ref{fig:Densitylargesystem} (b), (d)). Since amylose and amylopectin share identical chemical units, this interfacial redistribution on the addition of clay platelets is primarily entropically driven by more efficient interfacial packing of linear amylose chains. Sorbitol, the most polar TPS constituent, shows negligible interfacial accumulation in both TPS-PE and TPS-PE-CL systems. The latter contrasts sharply with sorbitol's strong adsorption to TMA-MMT clay (Figure \ref{fig:de} (d)), driven by replacement of polar TMA groups with nonpolar C12TAB grafts. Figure \ref{fig:sheetsnapshots} shows representative snapshots for organization of C12TAB-MMT in the PE bulk and at the interface in initially prepared, locally equilibrated (after T-P cycling), and final (at the end of \SI{1500}{\nano\second}) simulation systems, highlighting significant rearrangement made accessible by the use of CG forcefield. Overall, our results demonstrate the developed CG model's accuracy in prediction of surface polarity-driven component redistribution in clay-polymer composites, establishing its utility for rational system design.

\subsubsection{Conformational Properties}     
\begin{figure}[!htb]
     \centering
     \includegraphics[width=17.65cm,height=6cm]{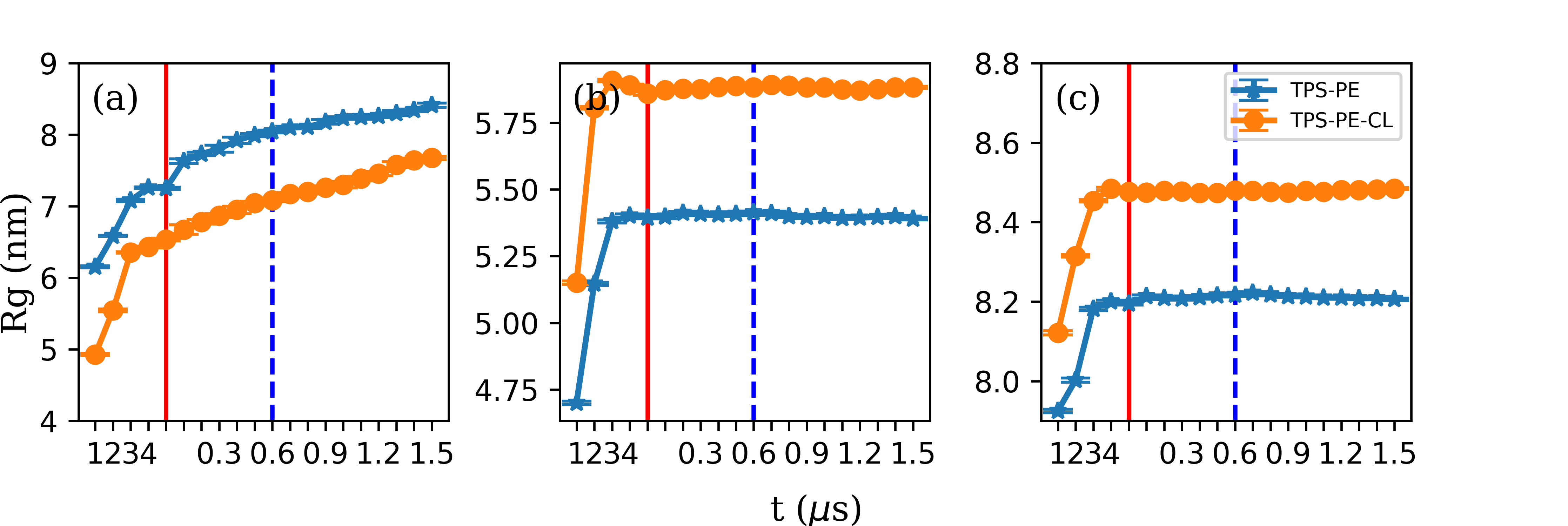}
     \caption{ \textbf{Dynamic changes in radius of gyration for polymers in TPS-PE melt and TPS-PE-CL composite system}. \textbf{(a)}, \textbf{(b)}, and \textbf{(c)} show the time evolution of radius of gyration ($R_\mathrm{g}$) of polyethylene (PE), amylose (AMY), and amylopectin (AMP), respectively. Data points calculated in \SI{100}{\nano\second} blocks from CG MARTINI simulations of TPS-PE melt ($\bluestar$) and TPS-PE-CL composite ($\orangewithcircle$) systems at \SI{613}{\kelvin} and \SI{1}{\bar}. The $x$-axis indicates the equilibration cycle number, a \SI{25}{\nano\second} $NVT$ simulation (indicated by a horizontal \redline), and the production run time scale in \SI{}{\micro\second}. This $\bluedotline$ represents the end of the equilibration period, and data points following this line were used to calculate the average and error values reported in Table \ref{tab:conformation}.}
     \label{fig:Rg}
 \end{figure}

The $R_\mathrm{g}$ of all polymers increased during annealing and initial MD simulation (Figure \ref{fig:Rg}), reaching a plateau in \SIrange{600}{900}{\nano\second}. Unfavorable polymer-clay interactions decreased PE $R_\mathrm{g}$ from \SI{8.24}{\nano\meter} in the melt to \SI{7.41}{\nano\meter} in the composite, while favourable interactions expended amylose and amylopectin chains from \SI{5.39}{\nano\meter} and \SI{8.21}{\nano\meter} to \SI{5.88}{\nano\meter} and \SI{8.48}{\nano\meter}, respectively. A 3x higher clay-TPS CN compared to clay-PE CN for interface clay (Figure \ref{fig:CN}) and clay tactoid formation in PE bulk confirms distinct clay-polymer interactions. Further, near- and far-region $R_\mathrm{g}$ ratio was 1.1 and 0.9 for amylose and PE, confirming chain expansion and contraction at the clay surface, respectively (Table \ref{tab:conformation}). However, this $R_\mathrm{g}$ ratio for amylopectin was 0.87, inspite of an overall expansion of amylopectin chain in the composite system. This apparent contradiction can be explained by preferential amylose-clay interactions leading to displacement of amylopectin and lowering of its coordination with the clay particle (Figure \ref{fig:Densitylargesystem}), and thus causing localized chain compaction in the near-surface region. This demonstrates that in multi-component systems, polymer conformations are governed by synergistic, collective interactions rather than individual pairwise interactions, emphasizing the role of competitive adsorption in determining morphology. Furthermore, Table \ref{tab:conformation} shows that the clay particle had a negligible effect on $S_\mathrm{c}$ of PE as its interaction with clay is primarily through chemically and structurally similar alkyl grafts, while amylose and amylopectin $S_\mathrm{c}$ decreased by \SIrange{3.5}{9}{\percent} in the composite system, consistent with a larger effect of clay on their chain conformation and near-surface packing as a result of polymer chain interactions with a rigid clay surface.

\begin{table*}
 \small
  \centering
  \captionof{table}{\textbf{Comparison of AA and CG properties in TPS-PE melt and TPS-PE-CL composite system.} The average values and error values for interfacial surface tension ($\gamma$), configuration entropy per CG bead ($S_\mathrm{c}$), and radius of gyration ($R_\mathrm{g}$) were calculated from the stable regions of the \SI{1500}{\nano\second} production runs, performed at \SI{613}{\kelvin} and \SI{1}{\bar}. These stable regions are highlighted by \bluedotline in Figure \ref{fig:surfacetension} ($\gamma$), \ref{fig:Sc} ($S_\mathrm{c}$), and \ref{fig:Rg} ($R_\mathrm{g}$), respectively.}
  \centering
  \label{tab:conformation}
\begin{threeparttable}
  
  \begin{tabular}{ c c c c c c c c c} 
    \toprule
 \multicolumn{1}{c}{} & \multicolumn{4}{c}{\textbf{TPS-PE}} &   \multicolumn{4}{c}{\textbf{TPS-PE-CL}} \\
\midrule
 \multicolumn{1}{c}{$\gamma$(\SI{}{\milli\newton\per\meter})} & \multicolumn{4}{c}{45.00$\pm$1.98} &   \multicolumn{4}{c}{29.03$\pm$5.41} \\
            \midrule
    
      \multicolumn{1}{c}{} & \multicolumn{2}{c} {\textbf{Amylose}} & &\multicolumn{2}{c}{\textbf{Amylopectin}}&&\multicolumn{2}{c}{\textbf{polyethylene}}   \\
    \cmidrule(lr){2-3}\cmidrule(lr){5-6}\cmidrule(lr){7-9}
     \multicolumn{1}{c}{} & \multicolumn{1}{c}{\textbf{TPS-PE}} &   \multicolumn{1}{c}{\textbf{TPS-PE-CL}}  &  & \multicolumn{1}{c}{\textbf{TPS-PE}} &  \multicolumn{1}{c}{\textbf{TPS-PE-CL}} & & \multicolumn{1}{c}{\textbf{TPS-PE}} &  \multicolumn{1}{c}{\textbf{TPS-PE-CL}} \\
    \midrule
    $S_\mathrm{c}$ $(\SI{}{\joule\per\mol\kelvin})$  &   24.21 $\pm$0.30    & 23.33 $\pm$0.10     &   & 24.28 $\pm$0.03 & 22.34 $\pm$0.12&&69.29 $\pm$0.07&69.64 $\pm$0.06 \\
      $R_\mathrm{g}$(\SI{}{\nano\meter})  &   5.30 $\pm$0.01    & 5.88 $\pm$0.01     &   & 8.21 $\pm$0.00 & 8.48 $\pm$0.00 &&8.24 $\pm$0.10&7.41 $\pm$0.18 \\
      \midrule
        \multicolumn{1}{c}{} & \multicolumn{2}{c}{\textbf{TPS-PE-CL}}  &  &   \multicolumn{2}{c}{\textbf{TPS-PE-CL}} & &  \multicolumn{2}{c}{\textbf{TPS-PE-CL}} \\
        \midrule
       &  near &  far&   & near &far && near&far  \\
           \cmidrule(lr){2-3}\cmidrule(lr){5-6}\cmidrule(lr){7-9}
       $R_\mathrm{g}$(\SI{}{\nano\meter})&   5.98 $\pm$0.09 &   5.46 $\pm$0.05&   &  7.89 $\pm$0.21 &  9.07 $\pm$0.24 && 6.54 $\pm$0.11&  7.26 $\pm$0.09\\  
        
                   \bottomrule
  \end{tabular}
  \end{threeparttable}
\end{table*}

\subsubsection{Interfacial Surface Tension}
We used the local pressure tensor at the TPS-PE interface to calculate the interfacial tension, $\gamma$, as the difference between its normal and tangential components. The normal component of the pressure tensor ($P_N$) was fixed by the use of $NP_zT$ ensemble, while significant fluctuations in the tangential component ($P_T$) indicate continuous interfacial rearrangement. These fluctuations arise from curvature effects and interfacial roughness, amplified by evolution of filler-polymer interactions, leading to convergence times of \SI{900}{\nano\second} in the composite and \SI{600}{\nano\second} in the melt. The average $\gamma$ for melt and composite systems was obtained to be 45 $\pm$ 1.98\SI{}{\milli\newton\per\meter} and 29 $\pm$ 5.41\SI{}{\milli\newton\per\meter}, respectively. The composite system features two distinct interfaces because of the use of periodic boundary conditions: one formed between TPS, PE, and clay components, and another between TPS and PE, the latter being the same as the TPS-PE system without the clay particles. We take the $\gamma$ for the composite system as an average over these two distinct interfaces and take the $\gamma$ for the melt as an estimate for the second interface in the composite. This yielded a $\gamma$ of \SI{13.06}{\milli\newton\per\meter} for the TPS-PE-Clay interface. Here, we ignored the effect of bulk PE phase clay tactoids, supported by no contacts between bulk clay and TPS phase throughout the simulation (Figure \ref{fig:CN} (g)).\\

The $\gamma$ value for the TPS-PE-Clay interface was notably lower than that of the TPS-PE, indicating that clay acts as a compatibilizer at the interface, improving interfacial adhesion between the TPS and PE phases. This is consistent with experimental studies on other OMMT in TPS-PE systems \cite{Sabetzadeh2016,sharif2011study}, where the addition of a similar oMMT clay (Cloisite-15A, clay with two tallow chain substitutions) reduced the domain size of TPS from \SI{500}{\nano\meter} to \SI{200}{\nano\meter} in TPS-PE composites, a phenomenon directly attributable to reduced interfacial tension at the polymer-clay interface. The developed CG model's ability to accurately predict microscopic properties and macroscopic behavior is expected to enable computationally efficient determination of structure-property relationships of polymer-clay nanocomposites and aid rational design of these systems. 

\section{Conclusions}
In this work, we performed all-atom (AA) molecular dynamics simulations to investigate the structure and dynamics of thermoplastic starch (TPS) near pyrophyllite and TMA-MMT surfaces. Replacing the lower-polarity pyrophyllite surface with a higher-polarity TMA-MMT surface increased the extent of near-surface layering in the water and sorbitol distributions, accentuating starch polymer displacement in the first clay solvation layer. Further, for near-surface polymers, switching from pyrophyllite to TMA-MMT reduced normalized diffusion coefficients  ($D_\mathrm{n}$) by \SI{23}{\percent} to \SI{30}{\percent} and $S_\mathrm{c}$ by \SI{4}{\percent} to \SI{8}{\percent}, with minimal change in $R_\mathrm{g}$. We conjecture that high sorbitol concentration near TMA-MMT weakens TMA-MMT–TPS attraction, causing only modest reductions in diffusion and conformational entropy. Properties from AA simulations were used to refine the initial MARTINI-3 CG bead parameters for pyrophyllite and TMA-MMT surfaces, which were based on hydration energies ($\Delta F_{hyd}$) calculated using Lifshitz theory applied to experimental surface tension data. We found that $\Delta F_{hyd}$-based $\ce{SC_6}$ bead assignment for surface $\ce{SiO_2}$ in pyrophyllite optimally reproduced AA structural and thermodynamic properties (RDFs, density profiles ($\rho(z)$), $R_\mathrm{g}$, $S_\mathrm{c}$) of the TPS-pyrophyllite system. However, TMA-MMT surface $\ce{SiO2}$ representation using $\ce{SC_6}$ bead failed to account for localized polarity induced by adjacent highly electronegative oxygen atoms (bonded to octahedral $\ce{Mg}$), which resulted in lower sorbitol density near the surface. The dual bead surface representation, high polarity $\ce{SP_1}$ beads for $Mg$-connected (one-third of total surface bead) and $\ce{SC_6}$ for $Al$-connected $\mathrm{SiO_2}$ groups, reproduced optimal values for AA structural and thermodynamic properties. Furthermore, $D_\mathrm{n}$ for organic solvents and TPS components was underestimated in CG simulations (w.r.t. AA) by $\sim$ 35-50 \SI{}{\percent} indicated a limitation of the MARTINI-3 FF in accurately capturing dynamics at the solid-fluid interface, but a high accuracy for several structural and thermodynamic properties is expected to allow wider utility. These CG models enabled investigation of TPS-PE melt and TPS-PE-CL composite interfaces with C12TAB-MMT clay at system sizes and time scales inaccessible to AA simulations, analogous to experimental TPS-PE/oMMT blend interfaces \cite{Sabetzadeh2016,sharif2011study}. The coordination number (CN) results indicated that clay compatibilizes TPS-PE by binding TPS through its bare polar surface and PE via alkyl-mediated interactions (Figure \ref{fig:CN}). This compatibilization reduced the interfacial surface tension from \SI{45}{\milli\newton\per\meter} in TPS-PE melt to \SI{13.06}{\milli\newton\per\meter} in TPS-PE-CL composite, consistent with the trend of experimental TPS droplet size reduction from \SI{500}{\nano\meter} to \SI{200}{\nano\meter} upon addition of a similar oMMT clay (Cloisite-15A, a clay with dual tallow chains) \cite{Sabetzadeh2016,sharif2011study}. High surface coverage and chemical similarity to PE gave comparable clay-clay and clay-PE CN at the interface. However, efficient packing of short alkyl grafts and weak quadrupolar interactions resulted in 3x higher clay-clay CN, forming bimodal clay assemblies in PE bulk. Further, High TPS-clay CN increased amylose and amylopectin $R_\mathrm{g}$ from melt to composite, with near-to-far $R_\mathrm{g}$ ratios showing expansion for amylose (1.10) and contraction for PE (0.90). However, preferential amylose-clay interactions displaced amylopectin from the interface, causing localized chain compaction (near-to-far ratio = 0.87). The high TPS-clay coordination, TPS-PE compatibilization, and clay aggregation in PE follow trends consistent with experimental observations on similar systems \cite{Sabetzadeh2016,sharif2011study}, with the CG model capturing essential interfacial properties and long-time dynamics for efficient compatibilizer design in polymer nanocomposites.
\begin{acknowledgement}
The author thanks the High-Performance Computing facility of IIT Delhi for providing the resources needed for this work.

\end{acknowledgement}
\bibliography{achemso-demo}
\section{Supplementary Material}
\beginsupplement
\subsection{Temperature-Pressure Annealing Cycles}\label{annealing}
\begin{figure*}[!htb]
     \centering
     \includegraphics[width=14cm,height=8cm]{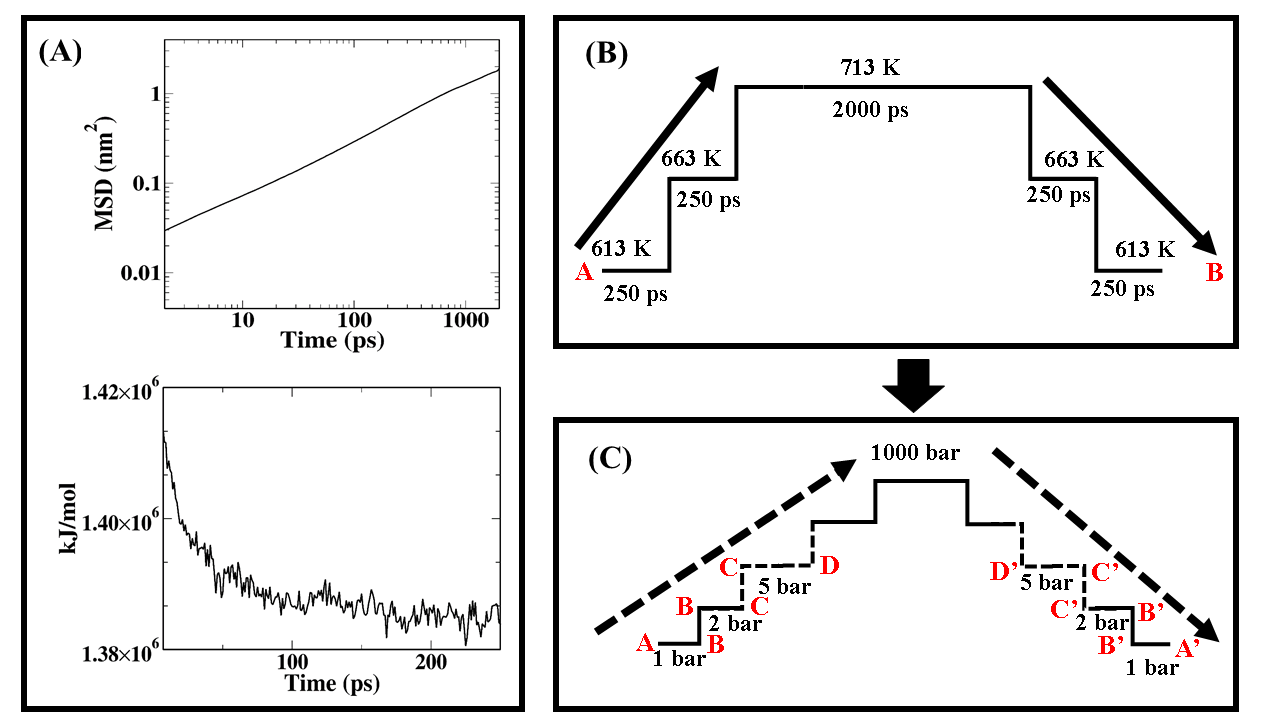}
     \caption{\textbf{Temperature-pressure annealing simulations}, The temperature cycle (\textbf{B}) was performed on the rescaled simulation box (state A). The figure shows the simulation time and temperature at each step. These temperature cycles were performed at various pressure values indicated in (\textbf{B}). The simulation time at higher (\SI{713}{\kelvin}) and lower (\SI{613}{\kelvin}) temperature is based mean-square displacement (at \SI{713}{\kelvin}) and potential energy (at \SI{613}{\kelvin}) (shown in \textbf{A}), respectively.} \label{fig:annealing}
 \end{figure*}

The initial box, obtained using Packmol software, underwent a multi-step density increase followed by multiple temperature-pressure (T-P) cycles to obtain an equilibrated TPS structure free from the memory effect of the initial configuration.
 The coordinates of the simulation box were rescaled in steps of \SI{29.17}{\cubic\nano\meter} (\SI{1}{\percent} of initial system volume) followed by energy minimization at each step using the steepest descent algorithm. The T-P annealing cycles were performed on the rescaled TPS system (referred to as "state A"). This was crucial for accelerating the slow dynamics inherent to starch chains, allowing the system to reach an equilibrated configuration more rapidly. The simulation time at the lower (\SI{613}{\kelvin}) and higher temperature (\SI{613}{\kelvin}) of the annealing cycle was chosen to ensure local equilibration (evidenced by a potential energy plateau) and Brownian relaxation of the polymer chain, respectively (Figure \ref{fig:annealing} (A)). Upon completion of each temperature cycle, the system pressure was systematically increased through successive steps with an approximate twofold increment per step, ultimately achieving a final pressure of \SI{1000}{\bar}. Once a pressure of \SI{1000}{\bar} was reached, the system was depressurized back to \SI{1}{\bar} using the same stepwise approach (Figure \ref{fig:annealing} (C)). This entire T-P cycling process was repeated until the densities of the initial (state A) and final (state A') states converged, confirming the attainment of an equilibrated structure (Figure \ref{fig:annealing} (C)). 
\subsection{Structures and Parameters}\label{structureandAAdetails}  
This section presents graphical representations of key systems and the bonded distributions for the sorbitol molecule.

\begin{figure}[!htb]
\centering
\includegraphics[width=8.35cm,height=10.20cm]{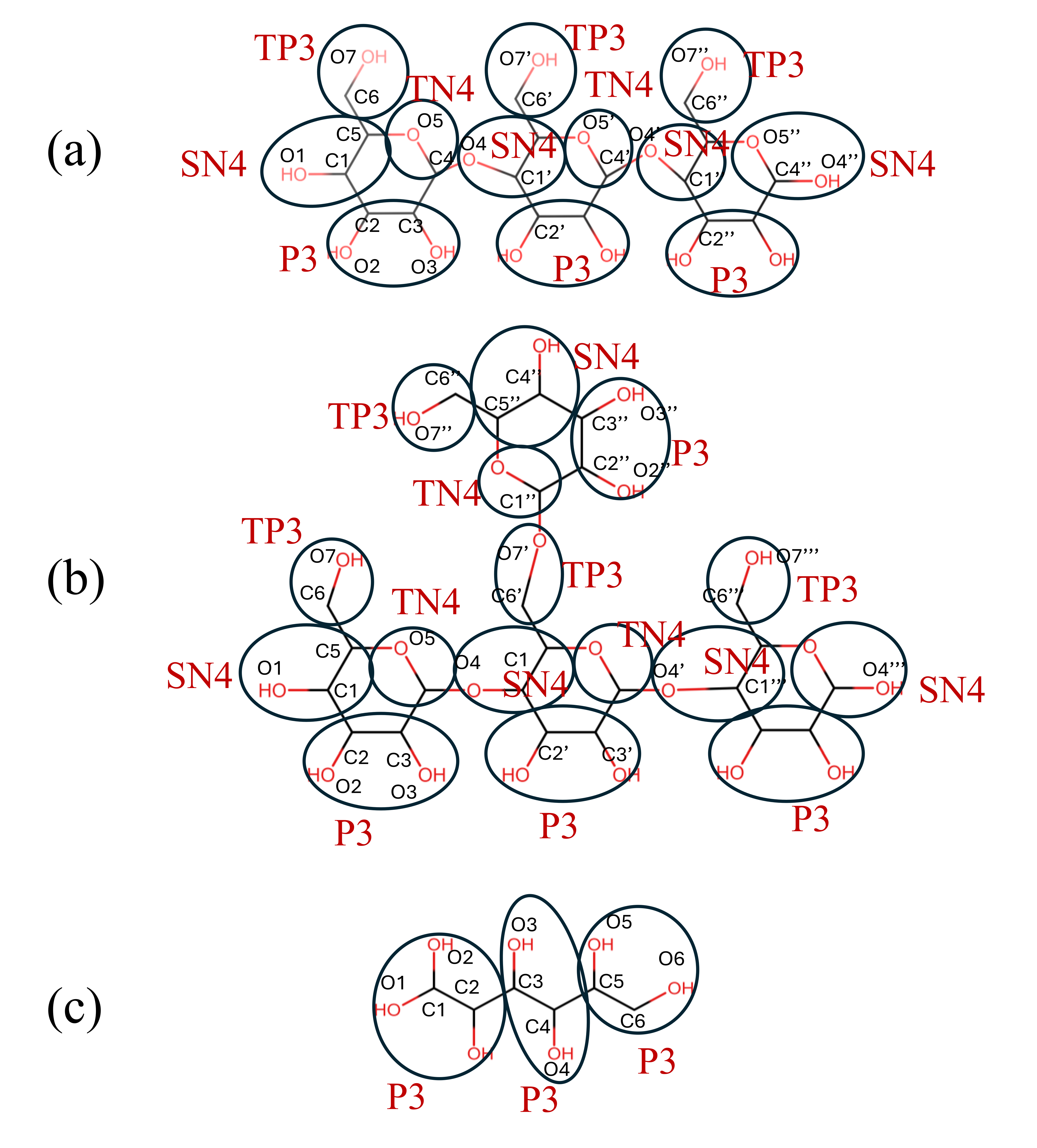}
\caption{ \textbf{CG MARTINI-3 mapping scheme for TPS.} Chemical representation for: \textbf{(a)} a trimeric segment of $\alpha$-D-glucose units connected with 1--4 linkage (representing a small unit of linear amylose chain), \textbf{(b)} 1--4 linked trimeric segment of $\alpha$-D-glucose with an additional 1--6 connection (represent branching in amylopectin), and \textbf{(c)} sorbitol. Atoms enclosed within circles are represented by a single CG bead, with MARTINI-3 bead type specified by red color.}
     \label{fig:mappingchemical}
 \end{figure}
\begin{table}[!htb]
  \small
  \caption{\textbf{Mapping Details for TPS.} The atom convention shown in Table is taken from Figure \ref{fig:mappingchemical}. The OH represents the hydrogen connected to oxygen.}
  \label{tab:mappingchemical}
  \scalebox{0.85}{
  \begin{tabular}{c c c}
    \toprule
    \textbf{Unit} & \textbf{CG Bead Type} & \textbf{Atoms} \\
    \midrule
    \textbf{1--4 Linked $\alpha$-D glucose} & & \\
    \midrule
    First monomer &TP3 & C6, O7, OH7 \\
    &TN4 & O5, C4, H4  \\
    &SN4 & C1, H1, O1, OH1, C5, H5 \\
    &P3 & C2, H2, O2, OH2, C3, H3, O3, OH3\\
    \midrule
    Middle monomer &TP3 & C6', O7', OH7' \\
    &TN4 &  O5', C4', H4' \\
    &SN4 & C1', H1', O4, C5', H5' \\
    &P3 & C2', H2', O2', OH2', C3', H3', O3', OH3'\\
    \midrule
    Terminal momomer&TP3 & C6", O7", OH7" \\
    &SN4 & O5", C4", H4", O4", OH" \\
    &SN4 & C1", H1", O4', C5", H5" \\
    &P3 & C2", H2", O2", OH2", C3", H3", O3", OH3"\\
    \midrule
    \textbf{1--6 Linked $\alpha$-D glucose branched monomers} & & \\
    \midrule
    Middle monomer &TP3 & C6', O7' \\
    &TN4 &  O5', C4', H4' \\
    &SN4 & C1', H1', O4, C5', H5' \\
    &P3 & C2', H2', O2', OH2', C3', H3', O3', OH3'\\
    \midrule
    Branched monomer&TP3 & C6", O7", OH7" \\
    &SN4 & C5", H5", C4", H4", O4", OH" \\
    &SN4 & C1", H1", O5"\\
    &P3 & C2", H2", O2", OH2", C3", H3", O3", OH3"\\
    \midrule
    \textbf{Sorbitol} & & \\
    &P3 & C1", H2", O2", OH2", C2", H2", O2", OH2"\\
    &P3 & C3", H3", O3", OH3", C4", H4", O4", OH4"\\
    &P3 & C5", H5", O5", OH5", C6", H6", O6", OH6"\\
    \midrule
    
    & \\
    \bottomrule
  \end{tabular}}
\end{table}
  
\subsection{Calculation of Surface Hydration Energies}\label{surface hydration energies}
The Lifshitz theory \cite{lifshitz1956theory} states that the surface tension ($\gamma$) consists of two components: the dispersive component ($\gamma^{LW}$) and the polar component ($\gamma^{AB}$), which account for dispersive and polar contributions to surface tension, respectively. The relationship between the total surface tension and its components is given by Equation \ref{eq:Lifshitzreal}.
\begin{equation}\label{eq:Lifshitzreal}
\gamma=\gamma^{LW}+\gamma^{AB} 
\end{equation}
The $\gamma^{AB}$ consists of donor ($\gamma^{-}$) and acceptor ($\gamma^{+}$) components, and relation can be expressed using Equation \ref{eq:polar}.
\begin{equation}\label{eq:polar}
\gamma_{l}^{AB} = 2\sqrt{ \gamma_{l}^{-}\gamma_{l}^{+}}
\end{equation}
The contact angle of liquids over a solid surface can be correlated with their surface tension components using Young's Equation \ref{eq:surfacetension} \cite{vanoss1988additive}. 
\begin{equation}\label{eq:surfacetension}
(1+cos\theta)\gamma_l = 2(\sqrt{ \gamma_{l}^{LW}\gamma_{s}^{LW}}) + \sqrt{ \gamma_{l}^{+}\gamma_{s}^{-}}+ \sqrt{ \gamma_{l}^{-}\gamma_{s}^{+}})
\end{equation}
Where $\gamma_{s}^{-}$, $\gamma_{s}^{+}$, $\gamma_{l}^{-}$, and $\gamma_{l}^{+}$ represent the donor and acceptor contributions to the solid and liquid polar components, respectively. The $\gamma_l$ is the surface tension of the liquid. This Equation \ref{eq:surfacetension}, along with reported surface tension data of polar and non-polar liquids (see Table \ref{tab:surface_energy}) \cite{giese1991surface}, was used for the calculation of $\gamma_{s}$.\\ 
In the first step, we have calculated the dispersive Lifshitz component ($\gamma_{s}^{LW}$) of pyrophyllite surface tension ($\gamma_{s}$) using Equation \ref{eq:surfacetension} for two non-polar liquids, i.e, $\alpha$-bromonaphthalene and diiodomethane. Since the polar contribution ($\gamma_{l}^{-}$ and $\gamma_{l}^{+}$) of these liquids are zero, we obtained $\gamma_{s}^{LW}$ of \SI{35.19}{\milli\newton\per\meter} for $\alpha$-bromonaphthalene and \SI{33.67}{\milli\newton\per\meter} for diiodomethane. The average of these two $\gamma_{s}^{LW}$ (\SI{34.4}{\milli\newton\per\meter}) was used for subsequent calculations.\\
\begin{table}[h!]
\scalebox{0.8}{
\centering
\begin{tabular}{|c|c|c|c|c|c|}
\hline
\textbf{Liquid} & \textbf{Contact Angle $\theta$} & $\boldsymbol{\gamma_{l}^{LW}}$ (\SI{}{\milli\newton\per\meter})& $\boldsymbol{\gamma_{l}^{AB}}$ (\SI{}{\milli\newton\per\meter}) & $\boldsymbol{\gamma_{l}^{-}}$ (\SI{}{\milli\newton\per\meter}) & $\boldsymbol{\gamma_{l}^{+}}$ (\SI{}{\milli\newton\per\meter}) \\
\hline
 $\alpha$-bromonaphthalene & \SI{38.7}{\degree} & 44.4 & 0 & 0 & 0 \\
\hline
 Diiodomethane & \SI{51.1}{\degree} & 50.8 & 0 & 0 & 0 \\
\hline
 Water & \SI{79.2}{\degree} & 21.8 & 51 & 25.5 & 25.5 \\
\hline
 Ethylene Glycol& \SI{48}{\degree} & 29.0 & 19.00 &47 & 1.92  \\
\hline
\end{tabular}}
\caption{\textbf{Surface energy components for different polarity liquids.} }
\label{tab:surface_energy}
\end{table}
In the second step, we have calculated the polar Lifshitz component ($\gamma_{s}^{-}$, and $\gamma_{s}^{+}$) of pyrophyllite surface tension ($\gamma_{s}$) using the Equation \ref{eq:surfacetension} over polar liquids, i.e., water and ethylene glycol. We obtained $\gamma_{s}^{+}$ of \SI{3.2}{\milli\newton\per\meter} and $\gamma_{s}^{-}$ of \SI{1.7}{\milli\newton\per\meter}.\\
In the third step, the values of $\gamma_{s}^{-}$, $\gamma_{s}^{+}$, and $\gamma_{s}^{LW}$ for pyrophyllite, as well as $\theta$, $\gamma_{l}^{-}$, $\gamma_{l}^{+}$, and $\gamma_{l}^{LW}$ for water, were integrated into Equation \ref{eq:hydration} to calculate the experimental hydration energy ($\Delta F_{hyd, exp}$) of the pyrophyllite surface. The calculation determined $\Delta F_{hyd, exp}$ as the difference between the free energy associated with solid particles interacting through a liquid medium ($f_{121}$) and the cohesive energy of solid particles in a vacuum ($f_{11}$). More details are provided in Section \ref{surface hydration energies derivation}.
\begin{equation}\label{eq:hydration}
\Delta F_{hyd, exp} = \mathrm{A_p}(-2 \gamma_{l}^{LW}+4\sqrt{ \gamma_{l}^{LW}\gamma_{s}^{LW}}) +(-2 \gamma_{l}^{AB}+4 \sqrt{ \gamma_{l}^{+}\gamma_{s}^{-}}+4\sqrt{ \gamma_{l}^{-}\gamma_{s}^{+}})
\end{equation}
The projected area ($\mathrm{A_p}=$ \SI{0.145}{\nano\meter\squared}) of the CG bead was used to obtain $\Delta F_{hyd, exp}$ of \SI{4.62}{\kilo\joule\per\mol}. However, it was reported that MARTINI-3 FF captures only relative trends rather than absolute hydration energy values. Therefore, we extrapolated the obtained $\Delta F_{hyd, exp}$ using Equation \ref{eq:extrapolationeq}. 
\begin{equation}\label{eq:extrapolationeq}
\Delta F_{hyd, sim}=0.65\Delta F_{hyd, exp}-4.52
\end{equation}

This relation is based on hydration energies obtained from experimental data and MARTINI simulations of thirty-six different organic liquids (reported in the recent MARTINI-3 study) \cite{souza2021martini}.\\
In the last step, the extrapolated hydration value (\SI{-1.65}{\kilo\joule\per\mol}) was directly compared against the hydration energies reported in the recent MARTINI publication \cite{souza2021martini}. We found that the $\Delta F_{hyd, sim}$ value of \SI{-1.65}{\kilo\joule\per\mol} is in close agreement with $\ce{SC_6}$ beads (\SI{-2.00}{\kilo\joule\per\mol}) \cite{souza2021martini}. Therefore, $\ce{SC_6}$ was chosen as the initial bead assignment for all surfaces of $\ce{SiO_2}$. 
\subsubsection{Hydration Energy Calculations}\label{surface hydration energies derivation}
Hydration energy can be expressed as the difference between two interaction energies, 
\begin{equation}
\Delta F_{hyd, exp} = f_{121} - f_{11}
\label{eq:refhydreq}
\end{equation}
where
\begin{itemize}
    \item $f_{121}$: Free energy when solid surfaces interact across liquid medium
    \item $f_{11}$: Cohesive energy when solid surfaces interact across vacuum
\end{itemize}
The cohesive energy can be written in terms of solid surface tension components using Equation \ref{eq:cohesive}.
\begin{equation}
f_{11} = -2\gamma_s = -2(\gamma_s^{LW} + \gamma_s^{AB})
\label{eq:cohesive}
\end{equation}

The $f_{121}$ can be written in terms of the solid-liquid interfacial tension using Equation \ref{eq: f121}.
\begin{equation}
f_{121} = -2\gamma_{sl}=2\gamma_s + 2\gamma_l - 4\sqrt{\gamma_s^{LW} \gamma_l^{LW}} - 4\sqrt{\gamma_s^+ \gamma_l^-} - 4\sqrt{\gamma_s^- \gamma_l^+}
\label{eq: f121}
\end{equation}

The substitution of $f_{121}$ and $f_{11}$ in Equation \ref{eq:refhydreq} will give $\Delta F_{hydration}$ in terms of surface tension components of solid (pyrophyllite) and liquid (water).  
\begin{equation}
\boxed{\Delta F_{hydration} = -2\gamma_l + 4\sqrt{\gamma_s^{LW} \gamma_l^{LW}} + 4\sqrt{\gamma_s^+ \gamma_l^-} + 4\sqrt{\gamma_s^- \gamma_l^+}}
\end{equation}
Here, the $\gamma_l$ is sum of $\gamma_l^{LW}$ and $\gamma_l^{AB}$ components.

\subsection{Key Results}  \label{key results}

\begin{table*}[!htb]
 \small
  \centering
  \captionof{table}{\textbf{Comparison of AA and CG normalized diffusion coefficients for five solvents in pyrophyllite-solvent systems} The normalized diffusion coefficients ($D_n$) for each solvent were calculated using different CG bead assignments for surface $\mathrm{SIO_2}$. All simulations were performed at \SI{300}{\kelvin} and \SI{1}{\bar}, except for the sorbitol system, which was simulated at \SI{450}{\kelvin} and \SI{1}{\bar}.}
  \centering
  \label{tab:solventDn}
\begin{threeparttable}
  
  \begin{tabular}{ c c c c c c c } 
    \toprule
    
    
    
\textbf{Solvent}& &Hexadecane  & 1-Propanol& Ethanol & Iso-Propanol & Sorbitol  \\
\cline{1-7}
 AA  & & 1.64$\pm$0.03  & 1.42$\pm$0.02 & 1.31$\pm$0.04  & 1.44$\pm$ 0.03& 1.62$\pm$0.02\\ 
\cline{1-7}
CG bead type &   & C6 &  N6   & SP1   & P1  & P4\\
\cline{1-7}
SC4 &   & 0.96 $\pm$0.04&  1.14 $\pm$0.06  & 0.96 $\pm$0.04  & 0.85 $\pm$0.03 & 0.93$\pm$0.02\\
SC6 &   & 1.04 $\pm$0.03&  1.09 $\pm$0.02 & 0.95 $\pm$0.05  & 0.82 $\pm$0.03 & 0.82$\pm$0.04\\
SN2 &   &  1.01 $\pm$0.04 &  1.06 $\pm$0.06  & 0.95 $\pm$0.02  & 0.80 $\pm$0.03 & 0.81$\pm$0.04\\
SN4 &   & 1.05 $\pm$0.02 & 1.03 $\pm$0.03 & 0.92 $\pm$0.02  & 0.77 $\pm$0.04 & 0.84$\pm$0.05\\
SN6 &   & 1.02 $\pm$0.05 &  1.06 $\pm$0.04  & 0.93 $\pm$0.04  & 0.79 $\pm$0.03 & 0.83 $\pm$0.02 \\
SP2 &   & 1.05 $\pm$0.03 &  1.07 $\pm$0.02  & 0.94$\pm$0.01  & 0.79 $\pm$0.03 & 0.84$\pm$0.06\\

 \bottomrule
  \end{tabular}
  \end{threeparttable}
\end{table*}

 \begin{equation}
   \mathrm{S_{2,\alpha\beta}}=-\pi\rho\int_{0}^{\infty}g_{\alpha\beta}(r)\mathrm{ln}\,g_{\alpha\beta} (r)-
   [g_{\alpha\beta}(r)-1]rdr
  \label{eqn:tbecomp}
\end{equation}

 \begin{equation}
    U(\xi) = \frac{1}{2} k \left(\xi - \xi_0\right)^2
    \label{eq:umbrella}
\end{equation}
 

  \begin{table*}[!htb]
 \small
  \caption{ \textbf{Properties of TPS melt in AA and CG simulations.} 
 All the properties of TPS melt were calculated at \SI{613}{\kelvin} and \SI{1}{bar}.} 
  \label{tab:tbemelt}
      \scalebox{0.8}{
      \begin{threeparttable}
  \begin{tabular}{c c c c c c c c c c}
    \toprule
      \multicolumn{2}{c}{System} &&& \multicolumn{3}{c}{$S_\mathrm{2}$ pairs} & \multicolumn{1}{c}{$\rho$ } (\SI{}{\kilo\gram\per\cubic\meter})&\multicolumn{2}{c}{$R_\mathrm{g}$ (\SI{}{\nano\meter})} \\
    \cmidrule(lr){1-10}
      \multicolumn{2}{c}{} & &&\multicolumn{1}{c}{Polymer-Polymer}  &  \multicolumn{1}{c}{Polymer-Sorbitol}  &  \multicolumn{1}{c}{Sorbitol-Sorbitol}&  \multicolumn{1}{c}{}& \multicolumn{1}{c}{Amylose}& \multicolumn{1}{c}{Amylopectin} \\
    \bottomrule
    
    AA &         &&& -1.33$\pm$0.02    & -0.36$\pm$0.01  & -0.52 $\pm$0.02&1214.69 $\pm$ 4.50 &1.38$\pm$ 0.02 &1.83$\pm$ 0.02\\
     CG MARTINI-2 &      &&&  -1.29 $\pm$0.03  & -0.75 $\pm$0.03 & -0.57 $\pm$0.02 &1136.00 $\pm$ 3.94 &1.58$\pm$ 0.05& 2.07$\pm$ 0.03\\
     CG MARTINI-3 &       &&& -1.30 $\pm$0.02   & -0.44 $\pm$0.02 & -0.56 $\pm$0.03 &1250.85 $\pm$ 3.23 &1.65$\pm$ 0.04&2.16$\pm$ 0.05 \\
    \hline
  \end{tabular}
\end{threeparttable}}
\end{table*}

The center of geometry positions of functional groups (corresponding to CG beads) in AA trajectories were converted to pseudo-CG trajectories using Equation \ref{eq:mapping}. 
\begin{equation}\label{eq:mapping}
\mathrm{r^{CG}_{i}}=\frac{\sum_{j=1}^{n}{r_{j}}}{n}
\end{equation}
Where, $\mathrm{r}_i$ and $n$ are positions and total number of atoms, $\mathrm{r^{CG}_{i}}$ are the positions for pseudo-CG beads. 

\begin{figure}[!htb]
     \centering
     \includegraphics[width=16.35cm,height=14.30cm]{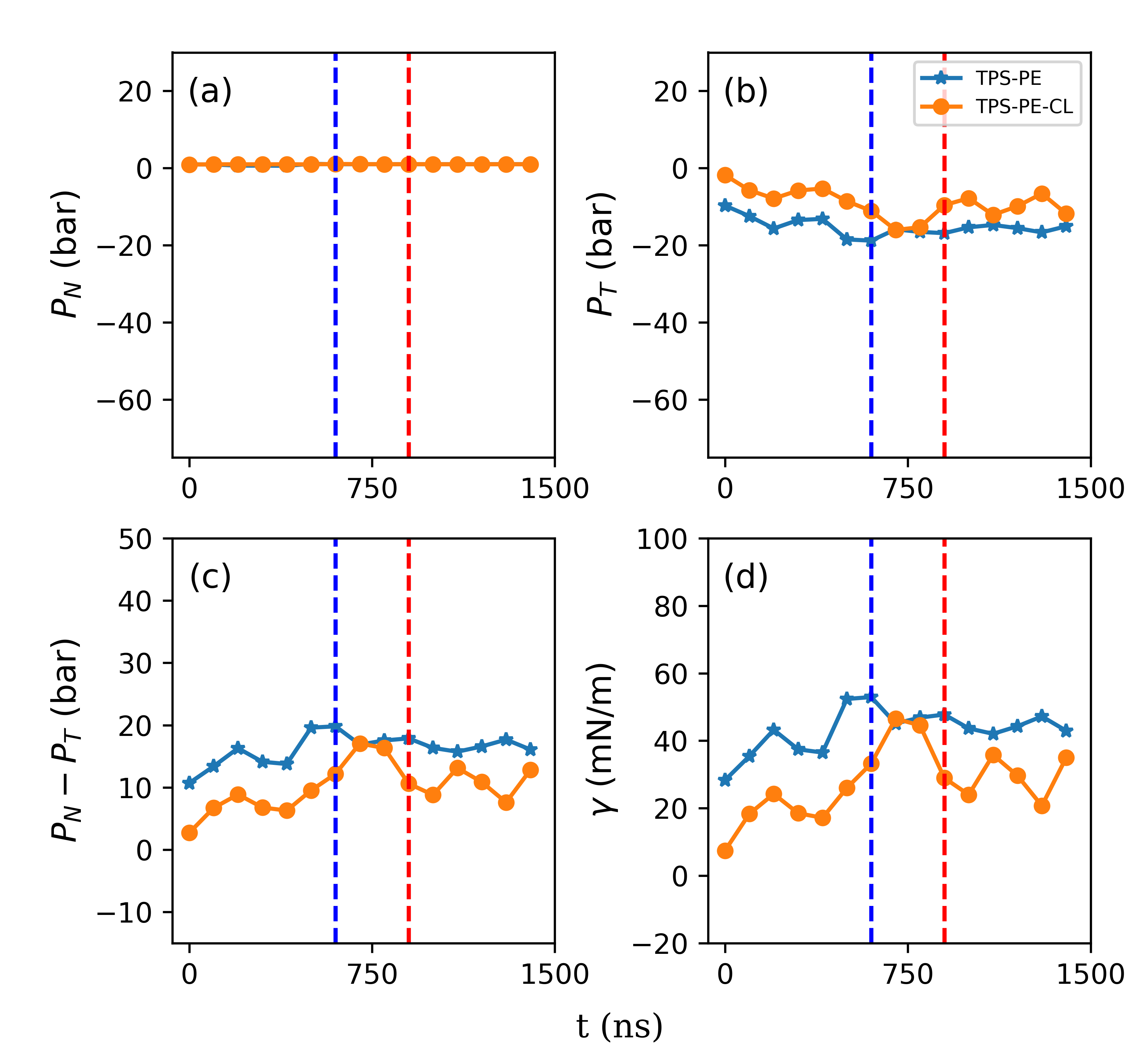}
     \caption{ \textbf{Time-dependent behavior of interfacial properties in TPS-PE melt and TPS-PE-CL composite systems.} This Figure presents the time evolution of: \textbf{(a)} The normal pressure component ($P_N$). \textbf{(b)} The tangential pressure ($P_T$) component, calculated as an average of $P_{xx}$ and $P_{yy}$. \textbf{(c)} The pressure anisotropy, determined by the difference between the normal and tangential components ($P_N-P_T$). \textbf{(d)} Interfacial surface tension ($\gamma$). The data obtained from CG simulation of TPS-PE melt (\bluestar) and TPS-PE-CL (\orangewithcircle) systems at \SI{613}{\kelvin} and \SI{1}{\bar}, with each data point showing average property over \SI{100}{\nano\second} blocks of \SI{1500}{\nano\second} trajectory.}
     \label{fig:surfacetension}
\end{figure}

\end{document}